\documentclass[journal]{IEEEtran}

\usepackage{cite}
\usepackage{algorithm}
\usepackage[noend]{algpseudocode}
\usepackage{amsmath,amssymb,amsfonts}
\usepackage{graphicx}
\usepackage{subcaption}
\usepackage{textcomp}
\usepackage{xcolor}
\usepackage{multirow}
\usepackage{makecell}
\usepackage{array}

\usepackage{pifont}

\begin{document}

\title{Semantic Communication Systems for Speech Transmission}

\author{Zhenzi Weng,~\IEEEmembership{Student Member,~IEEE,} and Zhijin Qin,~\IEEEmembership{Member,~IEEE} 
\thanks{Part of the work presented in\cite{Weng2106:Semantic}, which has been accepted by IEEE ICC 2021.}
\thanks{Zhenzi Weng and Zhijin Qin are with the School of Electronic Engineering and Computer Science, Queen Mary University of London, London E1 4NS, UK (email: zhenzi.weng@qmul.ac.uk, z.qin@qmul.ac.uk)(Corresponding author: Zhijin Qin).}
}

\maketitle

\begin{abstract}
Semantic communications could improve the transmission efficiency significantly by exploring the semantic information. In this paper, we make an effort to recover the transmitted speech signals in the semantic communication systems, which minimizes the error at the semantic level rather than the bit or symbol level. Particularly, we design a deep learning (DL)-enabled semantic communication system for speech signals, named DeepSC-S. In order to improve the recovery accuracy of speech signals, especially for the essential information, DeepSC-S is developed based on an attention mechanism by utilizing a squeeze-and-excitation (SE) network. The motivation behind the attention mechanism is to identify the essential speech information by providing higher weights to them when training the neural network. Moreover, in order to facilitate the proposed DeepSC-S for dynamic channel environments, we find a general model to cope with various channel conditions without retraining. Furthermore, we investigate DeepSC-S in telephone systems as well as multimedia transmission systems to verify the model adaptation in practice. The simulation results demonstrate that our proposed DeepSC-S outperforms the traditional communications in both cases in terms of the speech signals metrics, such as signal-to-distortion ration and perceptual evaluation of speech distortion. Besides, DeepSC-S is more robust to channel variations, especially in the low signal-to-noise (SNR) regime.

\end{abstract}

\begin{IEEEkeywords}
Deep learning, semantic communication, speech transmission, squeeze-and-excitation networks.
\end{IEEEkeywords}

\IEEEpeerreviewmaketitle

\section{Introduction}
\IEEEPARstart{I}{nspired} by the success in various areas, deep learning (DL) has been considered as a promising candidate for communications to achieve higher system performance with more intelligence\cite{qin2019deep,qin2020federated}. Particularly, DL has shown its great potentials to solve the existing technical problems in both physical layer communications\cite{gruber2017deep,ye2017power,8227772} and wireless resource allocations\cite{sun2018learning,liang2019deep}.

Typically, a DL-based communication system is designed to reduce the complexity and/or improve the system performance, by merging one or multiple communication modules in the traditional block-wise architecture and using a neural network (NN) to represent the intelligent transceiver. However, even if the DL-enabled communication systems yield better performance and/or lower complexity for some scenarios and conditions, their state-of-the-art models mainly focus on performance improvement at the bit or symbol level, which usually takes bit-error rate (BER) or symbol-error rate (SER) as the performance metric. Particularly, the major task in the traditional communication systems and the developed DL-enabled systems is to recover the transmitted message accurately and effectively, represented by digital bit sequences. In the past decades, such type of wireless communication systems have experienced significant development from the first generation (1G) to the fifth generation (5G) with the system capacity approaching Shannon limit. 

Shannon and Weaver\cite{weaver1953recent} categorized communications into three levels: 
\begin{itemize}
\item \emph{Level A}: how accurately can the symbols of communication be transmitted? (The technical problem)
\item \emph{Level B}: how precisely do the transmitted symbols convey the desired meaning? (The semantic problem)
\item \emph{Level C}: how effectively does the received meaning affect conduct in the desired way? (The effectiveness problem)
\end{itemize}
This indicates the feasibility to transmit the semantic information, instead of the bits or symbols, to achieve higher system efficiency. Besides, due to the wide deployment of intelligent IoT applications, semantic-irrelative communications are no longer ideal as they transmit bit sequences, which contains information that could be relevant or not to the intelligent tasks at the receiver. Moreover, in the typical communication systems, the generated data is more than required, which limits the number of devices to be covered by the same network. Motivated by this, researchers are dedicating to develop a new system to process and exchange semantic information for more efficient communications.

Semantic theory takes into account the meaning and veracity of source information because they can be both informative and factual\cite{carnap1952outline}, which facilitates the semantic communication systems to transmit only the semantic information at the transmitter and to recover information at the receiver via minimizing the semantic error instead of BER/SER. Nevertheless, the exploration of semantic communications has gone through decades of stagnation since it was first identified because of the limitation of some fundamental problems, e.g., lack of mathematical model for semantic information. The semantic information refers to the information relevant to the transmission goal at the receiver, however, even the most cutting-edge work cannot define the semantic information or semantic features by a precise mathematical formula. Moreover, the semantic information varies for different transmission purposes, which could be in various formats, e.g., age of information\cite{maatouk2020age}, or more complicated semantic features.

Semantic data can be compressed to a proper size for transmission by using a lossless method\cite{basu2014preserving}, which utilizes the semantic relationship between different messages, while the traditional lossless source coding is to represent a signal with the minimum number of binary bits by exploring the dependencies or statistical properties of input signals. In addition, an end-to-end (E2E) communication system has been developed\cite{o2017introduction} in order to address the bottlenecks in traditional block-wise communication systems. Inspired by this, different types of sources have been considered in recent investigations on E2E semantic communication systems, which mainly focuses on the image and text transmission\cite{guler2018semantic,9398576,xie2020lite,bourtsoulatze2019deep,kurka2020deepjscc,jankowski2020wireless,lee2019deep,jankowski2020joint}. The investigation on semantic communication for speech signals transmission is still missed.

\begin{figure}[tbp]
\includegraphics[width=0.45\textwidth]{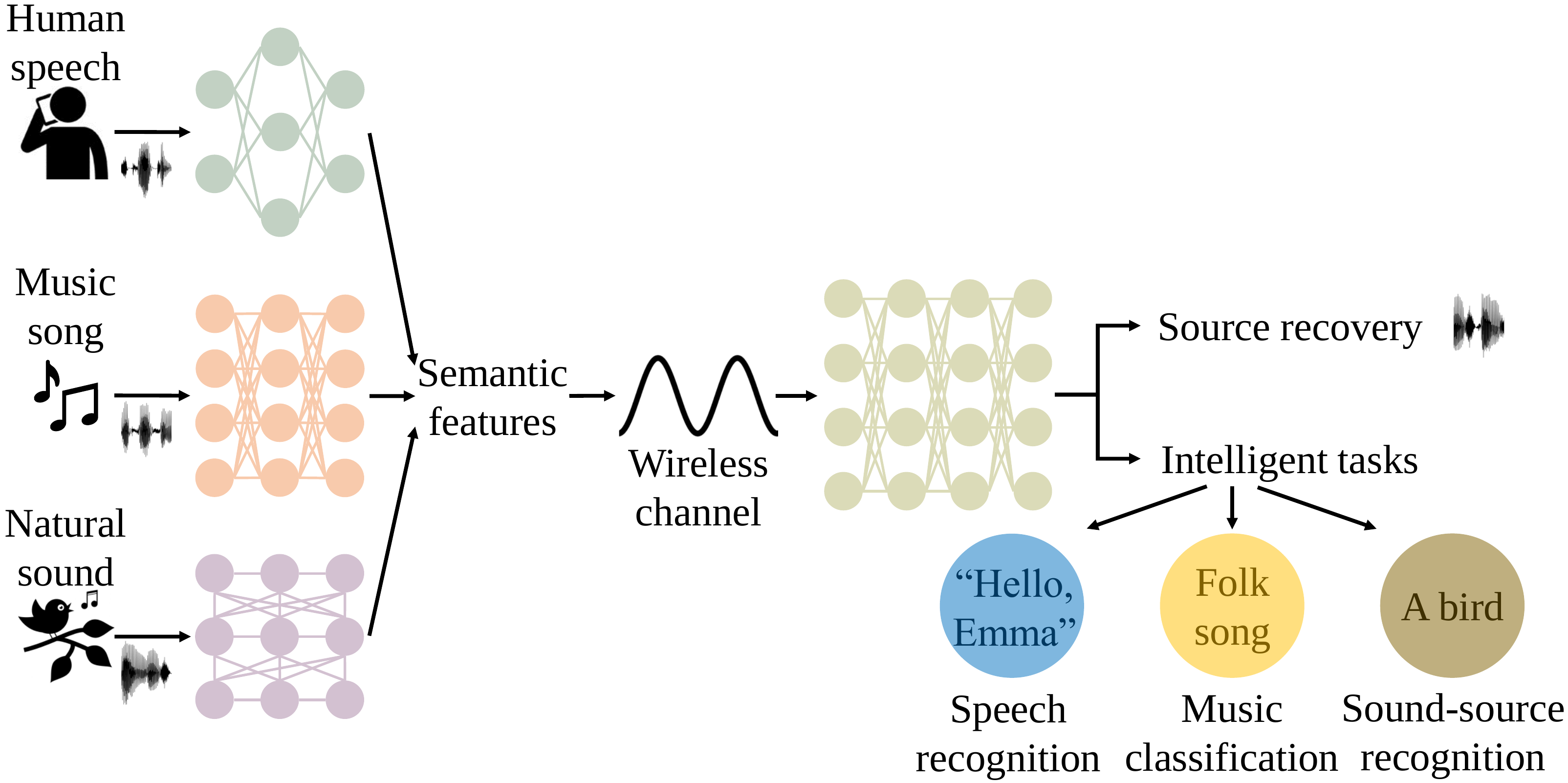} 
\centering 
\caption{A semantic communication system with different inputs.}  
\label{sound semantic communication}  
\end{figure}

A semantic communication system with different inputs is shown in Fig. \ref{sound semantic communication}, which only transmits the semantic features highly relevant to the transmission task at the receiver. While the transmission tasks could be either the source message recovery or more intelligent tasks. For example, in speech signal processing, one intelligent task is to convert speech signals into text information, e.g., automatic speech recognition (ASR), but do not care about the characteristics of speech signals, e.g., the speaking speed and tone\cite{dahl2011context}. The core of ASR is to map each phoneme into a single alphabet, and then concatenate all alphabets into an understandable word sequence via a language model. In this case, the extracted semantic features only contain the text characteristics while the other features will not be transmitted by the transmitter. As a result, the network traffic is reduced significantly without performance degradation.

Without loss of generality, we consider a general model of semantic communication to restore the speech sources in this work, which could be extended to serve a specific intelligent task easily in the future. The intuition behind this work is to recover speech signals based on DL technique, which includes the recovery of speech characteristics. However, most DL algorithms pre-process speech signals to obtain magnitude, spectra, or Mel-Frequency Cepstrum by various operations, such as discrete cosine transform (DCT), before feeding into a learning system. Such extra operations capture the unique features of speech signals but runs counter to the motivation of intelligence. Therefore, a DL-enabled semantic communication system by learning semantic information directly from the raw speech signals is of great interest and importance.

In this paper, we propose a DL-enabled semantic communication system for speech signals, named DeepSC-S, by learning and extracting speech signals, and then recovering them at the receiver from the received features directly. The main contributions of this article can be summarized as fourfold:

\begin{itemize}
\item The DeepSC-S is first developed, which treats the transmitter and the receiver as two NNs. The joint semantic-channel coding is developed to deal with source distortions and channel effects.

\item Particularly, a squeeze-and-excitation (SE) network\cite{8701503} is employed in the developed DeepSC-S to learn and extract essential speech semantic information, as well assign high values to the weights corresponding to the essential information during the training phase. By exploiting the attention mechanism based on an SE network, DeepSC-S improves the accuracy of signal recovery.

\item Moreover, a trained model with high robustness to channel variations is developed by training DeepSC-S under a fixed channel condition, and then facilitating it with good performance when coping with different testing channel environments.

\item To verify the model adaptation to practical communication scenarios, the proposed DeepSC-S is applied to telephone systems and multimedia transmission systems, respectively. The performance is also compared with the traditional approaches to prove its superiority, especially in the low SNR regime.

\end{itemize}

The rest of this article is structured as follows. The related work is presented in Section \uppercase\expandafter{\romannumeral2}. Section \uppercase\expandafter{\romannumeral3} introduces the model of semantic communication system for speech transmission and the related performance metrics. In Section \uppercase\expandafter{\romannumeral4}, the details of the proposed DeepSC-S are presented. Simulation results are discussed in Section \uppercase\expandafter{\romannumeral5}. Section \uppercase\expandafter{\romannumeral6} draws conclusions.

\emph{Notation}: The single boldface letters are used to represent vectors or matrices and single plain capital letters denote integers. Given a vector $\boldsymbol x$, $x_i$ indicates its $i$th component, $\left\|\boldsymbol x\right\|$ denotes its Euclidean norm. Given a matrix $\boldsymbol Y$, $\boldsymbol Y\in\mathfrak R^{M\times N}$ indicates that $\boldsymbol Y$ is a matrix with real values and its size is $M\times N$. Superscript swash letters refer the blocks in the system, e.g., $\mathcal T$ in $\boldsymbol\theta^{\mathcal T}$ represents the parameter at the transmitter and $\mathcal R$ in $\boldsymbol\theta^{\mathcal R}$ represents the parameter at the receiver. $\mathcal{CN}(\boldsymbol m,\;\boldsymbol V)$ denotes multivariate circular complex Gaussian distribution with mean vector $\boldsymbol m$ and co-variance matrix $\boldsymbol V$. Moreover, $\boldsymbol a\ast\boldsymbol b$ represents the convolution operation on the vector $\boldsymbol a$ and the vector $\boldsymbol b$.
\section{Related Work}
This section first introduce the related work on E2E communication systems to address the challenges in traditional systems. Then, we discuss the state-of-the-art semantic communications.

\subsection{End-to-End Communication Systems}
The DL-enabled E2E communication systems have achieved extremely competitive block-error rate (BLER) performance in comparison to the traditional block-wise communication systems in various scenarios\cite{o2017introduction}. In addition, it has shown great potentials in processing complicated communication tasks. For examples, the E2E learning has been employed in orthogonal frequency division multiplexing (OFDM) systems\cite{kim2017novel,felix2018ofdm}, as well as in multiple-input multiple-output (MIMO) systems\cite{o2017physical,he2020model}. Besides, channel estimation is a challenging problem in the DL-enabled E2E systems. In\cite{aoudia2019model}, reinforcement learning (RL) has been adopted to estimate the channel state information (CSI) through treating the channel layer and the receiver as the \emph{environment}. However, it requires a reliable channel to feedback the losses from the receiver to the transmitter during the training phase. Another novel channel agnostic solution has been proposed in\cite{ye2020deep}, which replaces the realistic channel with a deep neural network (DNN) by exploiting a conditional generative adversarial network (GAN).

Due to the complexity of NN training, a system with high training efficiency and low energy consumption is more than desired to make the E2E system applicable in practice. Transfer learning is a promising technology for adapting E2E communication systems to cope with the uncontrollable and unpredictable channel environments by training them over a statistical channel model\cite{pan2009survey}. Another appealing solution is to obtain a trained model yielding expected performance via the stochastic gradient descent (SGD) with a small number of iterations. Particularly, based on model-agnostic meta-learning (MAML), an E2E communication system has been developed in\cite{park2020meta}, which finds the initial NN parameters to achieve fast convergence for various channel conditions.

\subsection{Semantic Communications}
An initial research on semantic communication systems for text information was developed\cite{guler2018semantic}, which mitigates the semantic error by integrating the semantic inference and the physical layer communications to optimize the whole transceiver. However, such a text-based semantic communication system only measures the semantic error at the word level instead of the sentence level. Thus, a further investigation on semantic text transmission, named DeepSC, has been developed\cite{9398576} to deal with the semantic error at the sentence level with various length. Powered by the Transformer\cite{10.5555/3295222.3295349}, the semantic encoder and the channel encoder are co-designed to minimize the semantic error and to improve the system capacity. Moreover, the increasing deployment of smart IoT applications requires IoT devices to be capable of dealing with more complicated tasks, which runs counter to the limited computing capability of IoT devices. Inspired by this, a lite distributed semantic communication system for text transmission, named L-DeepSC\cite{xie2020lite}, has been further proposed to address the challenge of IoT devices by pruning and quantizing NN parameters. By doing so, the size of the trained model as well as the communication cost between the IoT devices and the server are reduced significantly, which makes it more suitable for IoT applications.

In semantic communications for image information, a DL-enabled semantic communication system has been developed\cite{bourtsoulatze2019deep}, which employs a convolutional neural network (CNN) at the transmitter to jointly design the source-channel coding. Besides, an image transmission system has been investigated to improve the accuracy of image reconstruction\cite{kurka2020deepjscc}, which backpropagates the channel output in order to generate a weight vector to the NN at the transmitter. In addition to perform the typical image reconstruction, more intelligent tasks have been considered. Particularly, an application based on a joint source-channel coding (JSCC) model to retrieve image has been proposed\cite{jankowski2020wireless}, which aims to reduce the transmission latency for IoT devices by achieving retrieval-oriented image compression. Besides, a joint image transmission-recognition system has been developed\cite{lee2019deep} for IoT applications, which has the superior recognition accuracy than the traditional approaches but requires the affordable computation resource. Moreover, by exploiting NN compression techniques, a deep JSCC\cite{jankowski2020joint} has been developed to perform image classification at the edge sever, which facilitates the IoT devices to process images with low computational complexity and to reduce the required transmission bandwidth. Note that the concept of semantic communications was not clearly stated by the authors in the aforementioned work on image transmission. However, we treat them as the pioneering works in the area as they share the spirit of semantic communications by extracting and transmitting the relevant information from the source for serving the transmission goals at the receiver.
\section{System Model}
In this section, we first introduce the considered system model, then the performance metrics are presented. The considered system transmits the original speech signals via a NN-based speech semantic communication system, which comprises two major tasks: \romannumeral1) semantic information learning and extracting from speech signals; \romannumeral2) mitigating the effects of wireless channels. For a practical communication scenario, the signal passing through the physical channel suffers from distortion and attenuation. Therefore, the considered DL-enabled system targets to recover the speech signals and to outperform the traditional approaches when coping with complicated channel conditions.

\subsection{Transmitter}\label{section system model-transmitter}
\begin{figure}[tbp]
\includegraphics[width=0.45\textwidth]{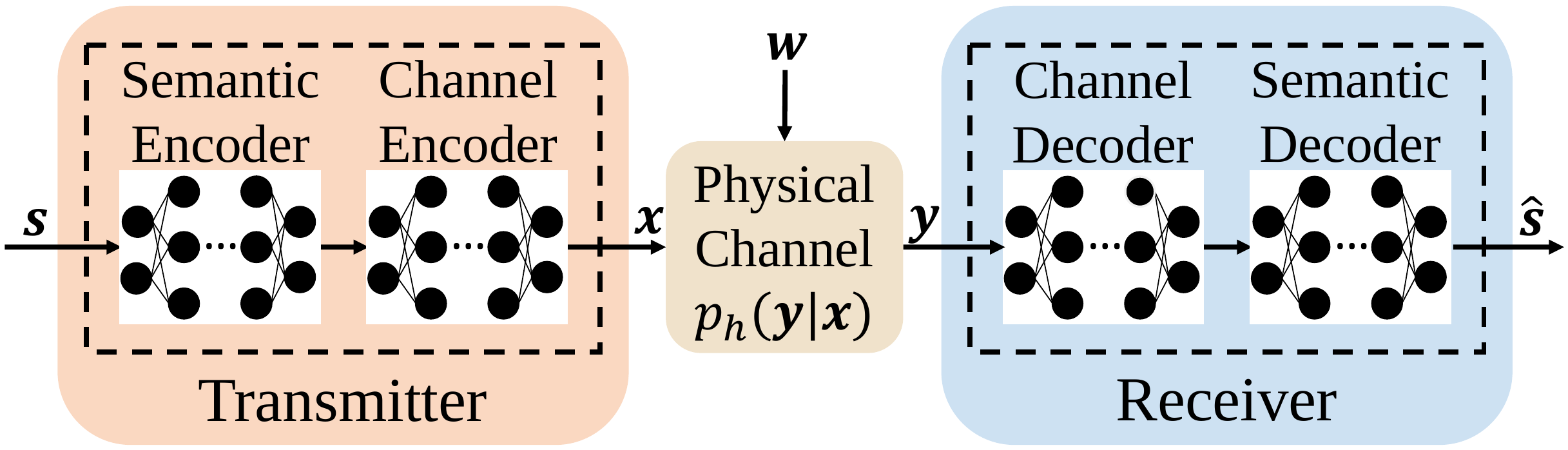} 
\centering 
\caption{The model structure of DL-enabled speech semantic communication system.}  
\label{sys model}  
\end{figure}

The proposed system model is shown in Fig. \ref{sys model}. From the figure, the input of the transmitter is a speech sample sequence, $\boldsymbol s=\left[s_1,\;s_2,\;...,\;s_W\right]$ with $W$ samples, where $s_w$ is $w$th item in $\boldsymbol s$ and it is a scalar value, i.e., a positive number, a negative number, or zero. At the transmitter, the input, $\boldsymbol s$, is mapped into symbols, $\boldsymbol x$, to be transmitted over physical channels. As shown in Fig. \ref{sys model}, the transmitter consists of two individual components: the \emph{semantic encoder} and the \emph{channel encoder}, each component is implemented by an independent NN. Denote the NN parameters of the \emph{semantic encoder} and the \emph{channel encoder} as $\boldsymbol\alpha$ and $\boldsymbol\beta$, respectively. Then the encoded symbol sequence, $\boldsymbol x$, can be expressed as
\begin{equation}
\boldsymbol x=\mathbf T_{\boldsymbol\beta}^{\mathcal C}(\mathbf T_{\boldsymbol\alpha}^{\mathcal S}(\boldsymbol s)),
\label{auto-encoder}
\end{equation}
where $\mathbf T_{\boldsymbol\alpha}^{\mathcal S}(\cdot)$ and $\mathbf T_{\boldsymbol\beta}^{\mathcal C}(\cdot)$ indicate the \emph{semantic encoder} and the \emph{channel encoder} with respect to (w.r.t.) parameters $\boldsymbol\alpha$ and $\boldsymbol\beta$, respectively. Here we denote the NN parameters of the transmitter as $\boldsymbol\theta^{\mathcal T}=(\boldsymbol\alpha,\boldsymbol\;\boldsymbol\beta)$.

The mapped symbols, $\boldsymbol x$, are transmitted over a physical channel. Note that the normalization on transmitted symbols $\boldsymbol x$ is required to ensure the total transmit power constraint $\mathbb{E}\left\|\boldsymbol x\right\|^2=1$.

The whole transceiver in Fig. \ref{sys model} is designed for a single communication link. The channel layer, represented by $p_h\left(\left.\boldsymbol y\right|\boldsymbol x\right)$, takes $\boldsymbol x$ as the input and produces the output as received signal $\boldsymbol y$. Denote the coefficients of a linear channel as $\boldsymbol h$, then the transmission process from the transmitter to the receiver can be modeled as
\begin{equation}
\boldsymbol y=\boldsymbol h\ast\boldsymbol x+\boldsymbol w,
\label{channel}
\end{equation}
where $\boldsymbol w\sim\mathcal{CN}(0,\;\sigma^2\mathbf I)$ indicates independent and identically distributed (i.i.d.) Gaussian noise, $\sigma^2$ is noise variance for each channel and $\mathbf I$ is the identity matrix.

\subsection{Receiver}
Similar to the transmitter, the receiver also consists of two cascaded parts, including the \emph{channel decoder} and the \emph{semantic decoder}. The \emph{channel decoder} is to mitigate the channel distortion and attenuation, and the \emph{semantic decoder} recovers speech signals based on the learned and extracted speech semantic features. Denote the NN parameters of the \emph{channel decoder} and the \emph{semantic decoder} as $\boldsymbol\chi$ and $\boldsymbol\delta$, respectively. As depicted in Fig. \ref{sys model}, the decoded signal, $\widehat{\boldsymbol s}$, can be obtained from the received signal, $\boldsymbol y$, by the following operation:
\begin{equation}
\widehat{\boldsymbol s}=\mathbf R_{\boldsymbol\delta}^{\mathcal S}(\mathbf R_{\boldsymbol\chi}^{\mathcal C}(\boldsymbol y)),
\label{auto-decoder}
\end{equation}
where $\mathbf R_{\boldsymbol\chi}^{\mathcal C}(\cdot)$ and $\mathbf R_{\boldsymbol\delta}^{\mathcal S}(\cdot)$ indicate the \emph{channel decoder} and the \emph{semantic decoder} w.r.t. parameters $\boldsymbol\chi$ and $\boldsymbol\delta$, respectively. Denote the NN parameters of the receiver as $\boldsymbol\theta^{\mathcal R}=(\boldsymbol\chi,\boldsymbol\;\boldsymbol\delta)$.

The objective of the whole transceiver system is to recover speech signals as close as to the original signals, which causes two challenges. The first one is the design of intelligent \emph{semantic encoder/decoder}, which utilizes the semantic information to recover speech signals, especially under the poor channel conditions, such as the low SNR regime. The second one is the design of the \emph{channel encoder/decoder} to alleviate symbol errors caused by the physical channels via adding redundancy information. For the traditional communications, the advanced channel coding techniques are achieved at the bit level to target a low BER. However, the bit-to-symbol transformation is not involved in our proposed system. The raw speech signals are directly mapped into a transmitted symbol stream by the \emph{semantic encoder} and the \emph{channel encoder}, and recovered at the receiver via inverse operations. Thus, we treat the speech recovery process as a signal reconstruction task to minimize the errors between the signal values in $\boldsymbol s$ and $\widehat{\boldsymbol s}$ by exploiting the characteristics of speech signals, then mean-squared error (MSE) is used as the loss function in our system to measure the difference between $\boldsymbol s$ and $\widehat{\boldsymbol s}$, denoted as
\begin{equation}
{\mathcal L}_{MSE}(\boldsymbol\theta^{\mathcal T},\;\boldsymbol\theta^{\mathcal R})=\frac1W\sum_{w=1}^W{(s_w-{\widehat s}_w)}^2,
\label{loss function}
\end{equation}
where $s_w$ and ${\widehat s}_w$ indicate the $w$th element of vectors $\boldsymbol s$ and $\widehat{\boldsymbol s}$, respectively. $W$ is the length of these two vectors.

Assume that the NN models of the whole transceiver are differentiable w.r.t. the corresponding parameters, which can be optimized via gradient descent based on (\ref{loss function}). It is worth to mention that the \emph{semantic encoder/decoder} and the \emph{channel encoder/decoder} are jointly designed. Besides, given prior CSI, both parameters sets $\boldsymbol\theta^{\mathcal T}$ and $\boldsymbol\theta^{\mathcal R}$ can be adjusted at the same time. Denote the NN parameters of the whole system model as $\boldsymbol\theta$, $\boldsymbol\theta=(\boldsymbol\theta^{\mathcal T},\boldsymbol\;\boldsymbol\theta^{\mathcal R})$, we adopt the SGD algorithm to train task in this paper, which iteratively updates the parameters $\boldsymbol\theta$ as follows:
\begin{equation}
\boldsymbol\theta^{(i+1)}\leftarrow\boldsymbol\theta^{(i)}-\eta\nabla_{\boldsymbol\theta^{(i)}}{\mathcal L}_{MSE}(\boldsymbol\theta^{\mathcal T},\;\boldsymbol\theta^{\mathcal R}),
\label{SGD}
\end{equation}
where $\eta>0$ is a learning rate and $\nabla$ indicates the differential operator.

\subsection{Performance Metrics}\label{section performance metric}
In our model, the system is committed to reconstruct the raw speech signals. Hence, the signal-to-distortion ration (SDR)\cite{vincent2006performance} is employed to measure the ${\mathcal L}_2$ error between $\boldsymbol s$ and $\widehat{\boldsymbol s}$, which is one of the commonly used metric for speech transmission and can be expressed as
\begin{equation}
SDR=10\log_{10}\left(\frac{\left\|\boldsymbol s\right\|^2}{\left\|\boldsymbol s-\widehat{\boldsymbol s}\right\|^2}\right).
\label{SDR}
\end{equation}

The higher SDR represents that the speech information is recovered with better quality, i.e., easier to understand for human beings. According to (\ref{loss function}), MSE loss could reflect the goodness of SDR. The lower the MSE, the higher the SDR.

Furthermore, the good recovery of speech signals is intuitively manifested by a satisfactory listening experience of the recovered speech signals, e.g., no latency and background noise. The perceptual evaluation of speech distortion (PESQ)\cite{rix2001perceptual} is adopted in the International Telecommunication Union (ITU-T) recommendation P.862\cite{ITU-T:p.862}, which is a good candidate for evaluating the quality of speech signals under various conditions, e.g., background noise, analog filtering, and variable delay, by scoring the quality from -0.5 to 4.5. The PESQ score is obtained by multiple operations, e.g., level align, time align and equalise, and disturbance processing\cite{rix2001perceptual}. In this work, we adopt an integrated open source PESQ assessment model developed by the ITU-T\cite{ITU-T:p.862}, which is able to evaluate the PESQ score in few milliseconds.
\section{Proposed Semantic Communication System for Speech Signals}
To address the aforementioned challenges, we design a DL-enabled semantic communication system for speech transmission, named DeepSC-S. Specifically, an attention-based two-dimension (2D) CNN is used for the \emph{semantic encoder/decoder} and a 2D CNN is adopted for the \emph{channel encoder/decoder}. The details of the developed DeepSC-S will be introduced in this section.

\subsection{Model Description}
\begin{figure*}[htbp]
\includegraphics[width=1\textwidth]{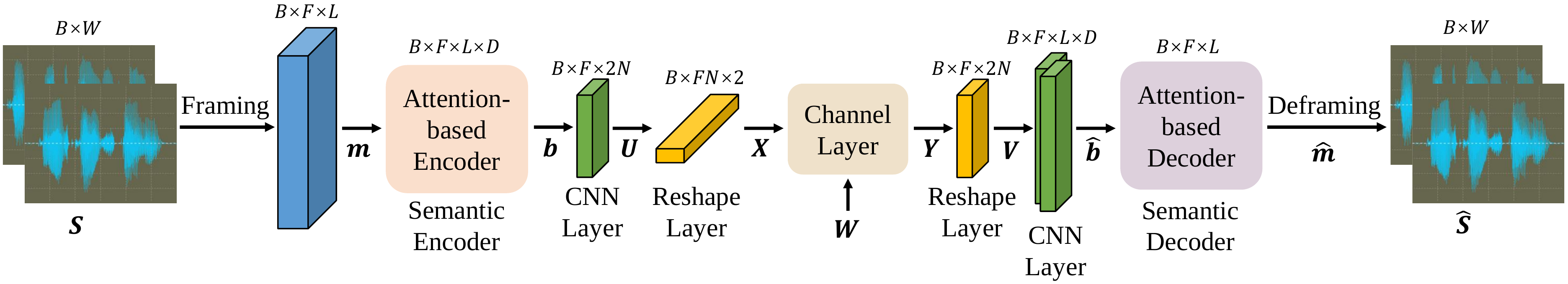} 
\centering 
\caption{The proposed system architecture for the speech semantic communication system.}  
\label{proposed sys}  
\end{figure*}

As shown in Fig. \ref{proposed sys}, the input of the proposed DeepSC-S, denoted as $\boldsymbol S\in\mathfrak R^{B\times W}$, is a set of speech sample sequences, $\boldsymbol s$, which is drawn from the speech dataset, $\mathfrak S$, and $B$ is the batch size. $\mathfrak S$ consists of considerable speech sequences, which are collected by recording the speakings from different persons. The input sample sequences, $\boldsymbol S$, are framed into $\boldsymbol m\in\mathfrak R^{B\times F\times L}$ for training before passing through an attention-based encoder, i.e., the \emph{semantic encoder}, where $F$ indicates the number of frames and $L$ is the length of each frame. Note that the framing operation only reshapes $\boldsymbol S$ without any feature learning and extracting. The \emph{semantic encoder} directly learns the speech semantic information from $\boldsymbol m$ and outputs the learned features $\boldsymbol b\in\mathfrak R^{B\times F\times L\times D}$. The details of the \emph{semantic encoder} are presented in part B of this section. Afterwards, the \emph{channel encoder}, denoted as a CNN layer with 2D CNN modules, converts $\boldsymbol b$ into $\boldsymbol U\in\mathfrak R^{B\times F\times2N}$. In order to transmit $\boldsymbol U$ into a physical channel, it is reshaped into symbol sequences, $\boldsymbol X\in\mathfrak R^{B\times FN\times2}$, via a reshape layer.

The channel layer takes the reshaped symbol sequences, $\boldsymbol X$, as the input and produces $\boldsymbol Y$ at the receiver, which is given by 
\begin{equation}
\boldsymbol Y=\boldsymbol H\boldsymbol X+\boldsymbol W,
\label{proposed channel}
\end{equation}
where $\boldsymbol H$ consists of $B$ number of channel coefficient vectors, $\boldsymbol h$, and $\boldsymbol W$ is Gaussian noise, which includes $B$ number of noise vectors, $\boldsymbol w$.

The received signal, $\boldsymbol Y$, is reshaped into $\boldsymbol V\in\mathfrak R^{B\times F\times2N}$ before feeding into the \emph{channel decoder}, represented by a CNN layer with 2D CNN modules. The output of the \emph{channel decoder} is $\widehat{\boldsymbol b}\in\mathfrak R^{B\times F\times L\times D}$. Afterwards, an attention-based decoder, i.e., the \emph{semantic decoder}, converts $\widehat{\boldsymbol b}$ into $\widehat{\boldsymbol m}\in\mathfrak R^{B\times F\times L}$ and $\widehat{\boldsymbol m}$ is recovered into $\widehat{\boldsymbol S}$ via the inverse operation of framing, named deframing. The size of $\widehat{\boldsymbol S}$ is same as that of $\boldsymbol S$ at the transmitter. The loss is calculated at the end of the receiver and backpropagated to the transmitter, thus, the trainable parameters in the whole system can be updated simultaneously.

\subsection{Semantic Encoder and Decoder}\label{section semantic encoder and decoder}
It is intuitive that speech signals at the silent time-slots carry no information, while speech signals at the speaking time-slots carry the essential information. Therefore, signals with zero magnitudes should be ignored, while signals with higher magnitudes should be processed with more attention. Moreover, at the speaking time-slots, people will speak loudly to emphasize the essential information. Correspondingly, the magnitudes of speech signals will increase abruptly. Similarly, people usually speak slowly to state the incomprehensible information, which results in a sudden drop on the frequency of corresponding speech signals. The essential information refers to the features that exist at the speaking time-slots, however, these features are hard or impossible to be captured by mathematical formula, which makes it difficult to measure the feature difference between the input speech signals and the recovered ones straightforwardly, or integrate them into loss function.

Inspired by this, we propose the DeepSC-S including the \emph{semantic encoder} and the \emph{semantic decoder} based on an attention mechanism, named SE-ResNet. In this work, SE-ResNet is utilized to learn and extract the essential information of speech signals. Particularly, SE-ResNet is capable to identify the essential information by assigning higher weights for them during the NN training process.

\begin{figure}[tbp]
\begin{minipage}[t]{1\linewidth}
\centering
\graphicspath{ {Figures/} } 
\includegraphics[width=0.9\textwidth]{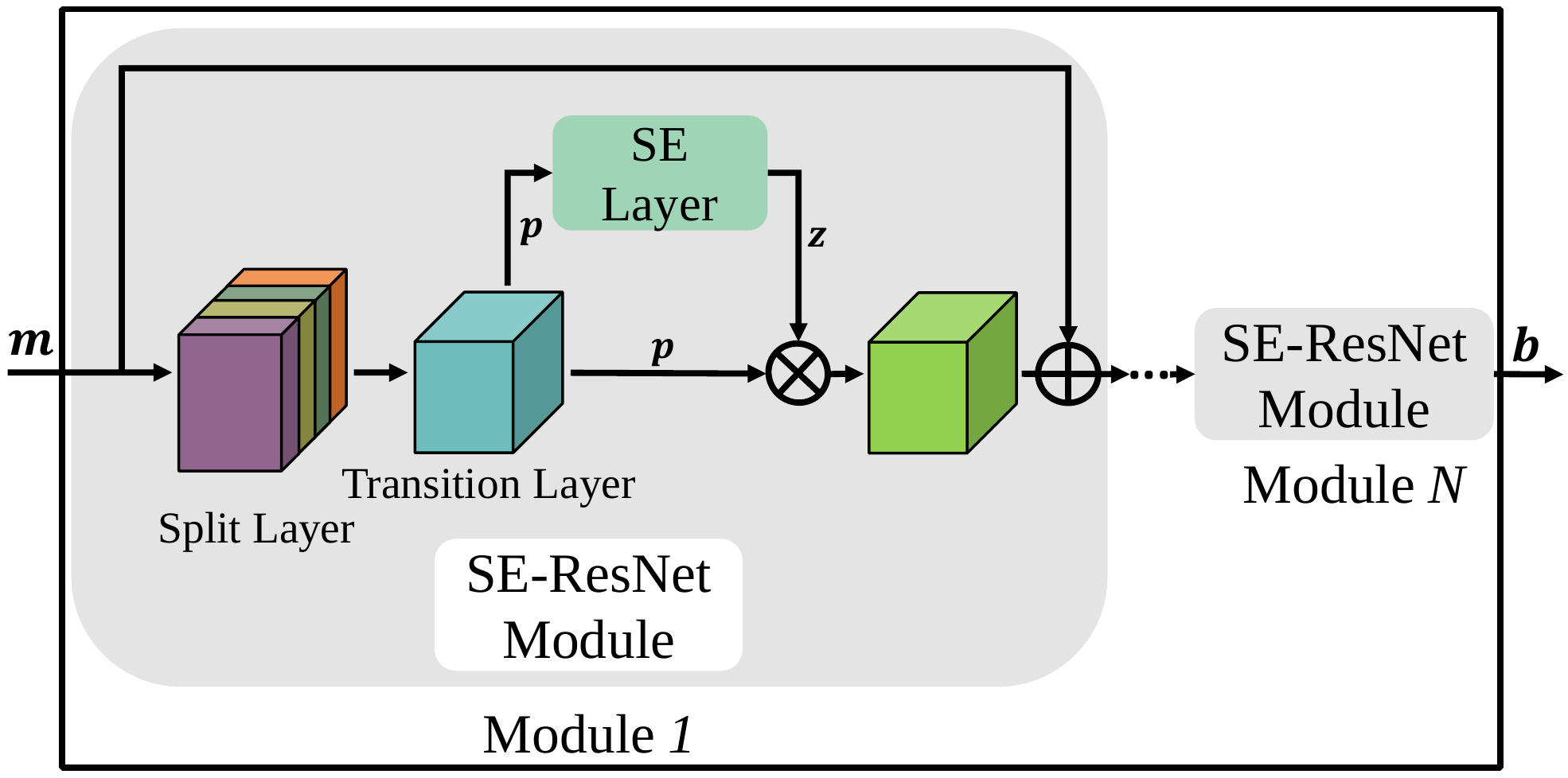}
\subcaption{Attention-based semantic encoder.}
\label{Attention-based Encoder}
\end{minipage}
\begin{minipage}[t]{1\linewidth}
\centering
\graphicspath{ {Figures/} } 
\includegraphics[width=0.9\textwidth]{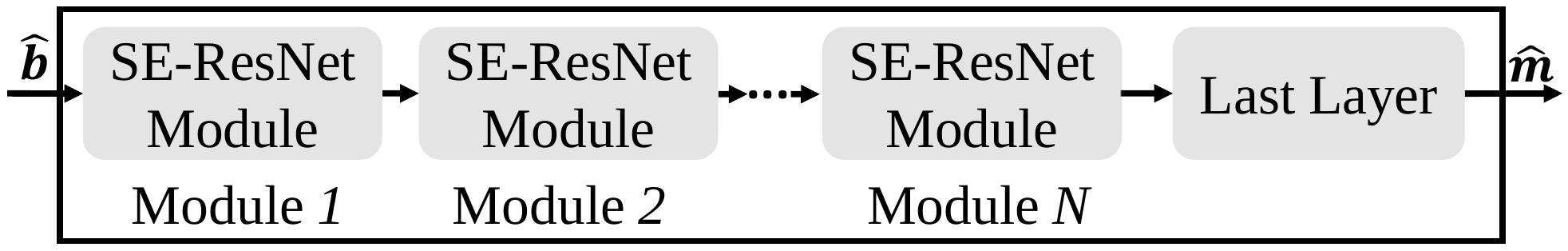}
\subcaption{Attention-based semantic decoder.}
\label{Attention-based Decoder}
\end{minipage} 
\caption{The proposed \emph{semantic encoder} and \emph{semantic decoder} based on SE-ResNet.}
\label{SE-ResNet}
\end{figure}

As shown in Fig. \ref{SE-ResNet}, for the SE-ResNet, a \emph{split} layer takes $\boldsymbol m$ as the input and produces multiples blocks, besides, all these blocks are concatenated. Then a \emph{transition} layer is utilized to reduce the dimension of these concatenated blocks and the output is denoted as $\boldsymbol p\in\mathfrak R^{M\times N\times C}$, which consists of $C$ features and each feature is in size of $M\times N$. For the SE layer, a \emph{squeeze} operation is employed to aggregate the 2D spatial dimension of each input feature, then an operation, named \emph{excitation}, intents to output the attention factor of each feature by learning the inter-dependencies of features in $\boldsymbol p$. The output of the SE layer, $\boldsymbol z\in\mathfrak R^{1\times1\times C}$, includes $C$ number of scale coefficients, which is utilized to scale the importance of the extracted features in $\boldsymbol p$, i.e., the features corresponding to the essential information in $\boldsymbol p$ are multiplied by the high scale coefficients in $\boldsymbol z$. By doing so, the weights of $\boldsymbol m$ are reassigned, i.e., the weights corresponding to the essential speech information are paid more attention. Note that the SE layer is an independent unit and one or multiple SE-ResNet modules can be sequentially connected. With more SE-ResNet modules, the performance of feature learning and extracting to the essential information will improve, however, it also increases the computational cost. Therefore, a trade-off between the learning performance and complexity should be considered during the training phase. Additionally, residual network is adopted to alleviate the problem of gradient vanishing due to the network depth by adding $\boldsymbol m$ into the output of SE-ResNet module, as shown in Fig. \ref{SE-ResNet}.

Particularly, the \emph{semantic encoder} is comprised by multiple SE-ResNet modules to convert input $\boldsymbol m$ into $\boldsymbol b$, corresponding to Fig. \ref{proposed sys}. For the \emph{semantic decoder}, in addition to several SE-ResNet modules, the \emph{last layer}, including a 2D CNN module with a single \emph{filter}, is utilized to reduce the size of output $\widehat{\boldsymbol m}$, because the size of $\boldsymbol m$ and $\widehat{\boldsymbol m}$ should be equal.

\subsection{Model Training and Testing}
Based on the prior knowledge of CSI, the transmitter and the receiver parameters, $\boldsymbol\theta^{\mathcal T}$ and $\boldsymbol\theta^{\mathcal R}$, can be updated simultaneously. As aforementioned, the objective of the proposed DeepSC-S is to train a model to recover the speech signals and make it to work well under various fading channels and a wide SNR regime.

\subsubsection{Training Stage} 
According to Fig. \ref{proposed sys}, the training algorithm of DeepSC-S is described in Algorithm \ref{training algorithm}. During the training stage, in order to facilitate the fast MSE loss convergence, the NN parameters, $\boldsymbol\theta=(\boldsymbol\theta^{\mathcal T},\boldsymbol\;\boldsymbol\theta^{\mathcal R})$, are initialized by a variance scaling initializer instead of 0. Besides, for achieving a valid training task, the MSE loss converges until the loss is no longer decreasing. The number of SE-ResNet modules is an important hyperparameter, which aims to facilitate the good performance of the \emph{semantic encoder/decoder} and the reasonable training time. Moreover, noise $\boldsymbol W$ in the channel layer is generated by a fixed SNR value.

After passing through the whole transceiver, the sample sequences set, $\boldsymbol S$, is recovered into $\widehat{\boldsymbol S}$, the size of $\boldsymbol S$ and $\widehat{\boldsymbol S}$ should be equal. Furthermore, the loss is computed at the end of the receiver according to (\ref{loss function}) and the parameters are updated by (\ref{SGD}).
\begin{algorithm}[htb]
\caption{Training algorithm of the proposed DeepSC-S}
\label{training algorithm}
\textbf{Initialization:} initialize parameters $\boldsymbol\theta^{\mathcal T{\boldsymbol(\mathbf0\boldsymbol)}}$ and $\boldsymbol\theta^{\mathcal R{\boldsymbol(\mathbf0\boldsymbol)}}$, $i=0$.

\begin{algorithmic}[1]
    \State \textbf{Input:} Speech sample sequences $\boldsymbol S$ from speech dataset $\mathfrak S$, fading channel $\boldsymbol H$, noise $\boldsymbol W$ generated under a fixed SNR value.
    \State Framing $\boldsymbol S$ into $\boldsymbol m$ with trainable size.
        \While{Stop criterion is not meet}
            \State $\mathbf T_{\boldsymbol\alpha}^{\mathcal S}(\boldsymbol m)\rightarrow\boldsymbol b$.
            \State $\mathbf T_{\boldsymbol\beta}^{\mathcal C}(\boldsymbol b)\rightarrow\boldsymbol X$.
            \State Transmit $\boldsymbol X$ over physical channel and receive $\boldsymbol Y$ via \State (\ref{channel}).
            \State $\mathbf R_{\boldsymbol\chi}^{\mathcal C}(\boldsymbol Y)\rightarrow\widehat{\boldsymbol b}$.
            \State $\mathbf R_{\boldsymbol\delta}^{\mathcal S}(\widehat{\boldsymbol b})\rightarrow\widehat{\boldsymbol m}$.
            \State Deframing $\widehat{\boldsymbol m}$ into $\widehat{\boldsymbol S}$.
            \State Compute loss ${\mathcal L}_{MSE}(\boldsymbol\theta^{\mathcal T},\;\boldsymbol\theta^{\mathcal R})$ via (\ref{loss function}).
            \State Update trainable parameters simultaneously via SGD:
                \begin{equation}
                \boldsymbol\theta^{\mathcal T\left(i+1\right)}\leftarrow\boldsymbol\theta^{\mathcal T\left(i\right)}-\eta\nabla_{\boldsymbol\theta^{\mathcal T\left(i\right)}}{\mathcal L}_{MSE}(\boldsymbol\theta^{\mathcal T},\;\boldsymbol\theta^{\mathcal R})
                \label{transmitter paramter update}
                \end{equation}
                \begin{equation}
                \boldsymbol\theta^{\mathcal R\left(i+1\right)}\leftarrow\boldsymbol\theta^{\mathcal R\left(i\right)}-\eta\nabla_{\boldsymbol\theta^{\mathcal R\left(i\right)}}{\mathcal L}_{MSE}(\boldsymbol\theta^{\mathcal T},\;\boldsymbol\theta^{\mathcal R})
                \label{receiver paramter update}
                \end{equation}
            \State $i\leftarrow i+1$.
        \EndWhile
    \State \textbf{end while}
    \State \textbf{Output:} Trained networks $\mathbf T_{\boldsymbol\alpha}^{\mathcal S}(\cdot)$, $\mathbf T_{\boldsymbol\beta}^{\mathcal C}(\cdot)$, $\mathbf R_{\boldsymbol\chi}^{\mathcal C}(\cdot)$, and $\mathbf R_{\boldsymbol\delta}^{\mathcal S}(\cdot)$.
\end{algorithmic}

\end{algorithm}

\subsubsection{Testing Stage}
Based on the trained networks $\mathbf T_{\boldsymbol\alpha}^{\mathcal S}(\cdot)$, $\mathbf T_{\boldsymbol\beta}^{\mathcal C}(\cdot)$, $\mathbf R_{\boldsymbol\chi}^{\mathcal C}(\cdot)$, and $\mathbf R_{\boldsymbol\delta}^{\mathcal S}(\cdot)$ from the outputs of Algorithm \ref{training algorithm}, the testing algorithm of DeepSC-S is illustrated in Algorithm \ref{testing algorithm}. Note that the speech sample sequences used for testing are different from that used for training.

The model is trained under a certain fading channel and a fixed SNR value, however, it is impractical to retrain the model for each possible channel condition and load all these models to the transmitter and the receiver. By comparing the testing results for models trained under various channel conditions, we adopt one of these models as the robust model, which could achieve good performance when coping with different channel environments. Accordingly, the robust model is employed to test the performance under various fading channels and different SNR values during the testing stage, as shown in Algorithm \ref{testing algorithm}.
\begin{algorithm}[htb]
\caption{Testing algorithm of the proposed DeepSC-S}
\label{testing algorithm}

\begin{algorithmic}[1]   
    \State \textbf{Input:} Speech sample sequences $\boldsymbol S$ from speech dataset $\mathfrak S$, trained networks $\mathbf T_{\boldsymbol\alpha}^{\mathcal S}(\cdot)$, $\mathbf T_{\boldsymbol\beta}^{\mathcal C}(\cdot)$, $\mathbf R_{\boldsymbol\chi}^{\mathcal C}(\cdot)$, and $\mathbf R_{\boldsymbol\delta}^{\mathcal S}(\cdot)$, testing channel set $\mathcal H$, a wide range of SNR values.
    \State Framing $\boldsymbol S$ into $\boldsymbol m$ with trainable size.
    	\For{each channel condition $\boldsymbol H$ drawn from $\mathcal H$}
    	    \For{each SNR value}
    	        \State Generate Gaussian noise $\boldsymbol W$ under the SNR value.
    	        \State $\mathbf T_{\boldsymbol\alpha}^{\mathcal S}(\boldsymbol m)\rightarrow\boldsymbol b$.
                \State $\mathbf T_{\boldsymbol\beta}^{\mathcal C}(\boldsymbol b)\rightarrow\boldsymbol X$.
                \State Transmit $\boldsymbol X$ over physical channel and receive $\boldsymbol Y$ \State via (\ref{channel}).
                \State $\mathbf R_{\boldsymbol\chi}^{\mathcal C}(\boldsymbol Y)\rightarrow\widehat{\boldsymbol b}$.
                \State $\mathbf R_{\boldsymbol\delta}^{\mathcal S}(\widehat{\boldsymbol b})\rightarrow\widehat{\boldsymbol m}$.
                \State Deframing $\widehat{\boldsymbol m}$ into $\widehat{\boldsymbol S}$.
                \EndFor
            \State \textbf{end for}
        \EndFor
    \State \textbf{end for}
	\State \textbf{Output:} Recovered speech sample sequences, $\widehat{\boldsymbol S}$, under different fading channels and various SNR values.
\end{algorithmic}

\end{algorithm}
\section{Experiment and Numerical Results}
In this section, we compare the performance of the proposed DeepSC-S, the traditional communication system, and the system with an extra feature coding for speech transmission under the additive white Gaussian noise (AWGN), Rayleigh, and Rician channels, where the accurate CSI is assumed. The details of the adopted benchmarks will be introduced in part A of this section. Moreover, in order to facilitate DeepSC-S with good adaptation to practical environments, it is tested over telephone systems and multimedia transmission systems. 

In the whole experiment, we adopt the speech dataset from Edinburgh DataShare, which comprises more than 10,000 \emph{.wav} files trainset and 800 \emph{.wav} files testset with sampling rate 16KHz. In terms of the traditional telephone systems and multimedia transmission systems, the sampling rates for speech signals are 8KHz and 44.1KHz, respectively. Thus, for the experiment regarding telephone systems, the input samples are down-sampled to 8KHz and regarding multimedia transmission systems, the input samples are up-sampled to 44.1KHz. Note that the number of speech samples in different \emph{.wav} is inconsistent. In the simulation, we fix $W=16,384,$ and each sample sequence in $\boldsymbol S$ consists of frames $F=128$ with the frame length $L=128$.

\subsection{Neural Network Setting and Benchmarks}
In the proposed DeepSC-S, the number of SE-ResNet modules in the \emph{semantic encoder/decoder} is 6. For each SE-ResNet module, the number of blocks in the \emph{split} layer is 2, which is achieved by 2 CNN modules with 16 \emph{filters} in each module, and the \emph{transition} layer is implemented by a CNN module with 32 \emph{filters}. Moreover, a single CNN module is utilized in the \emph{channel encoder/decoder}, which contains 8 \emph{filters}. The learning rate is 0.001. The parameters settings of the proposed DeepSC-S in telephone systems are summarized in Table \ref{telephone DeepSC-S NN parameters}. For performance comparison, we provide the following three benchmarks.

\subsubsection{\textbf{Benchmark 1}}
The first benchmark includes the typical source and channel coding techniques. According to ITU-T G.711 standard, 64 Kbps pulse code modulation (PCM) is recommended for speech source coding in telephone systems with $2^8=256$ quantization levels\cite{cox1997three}. Moreover, 16-bits PCM is adopted in our work for speech transmission in multimedia transmission systems with $2^{16}=65,536$ quantization levels. The A-law PCM and uniform PCM are adopted in telephone systems and multimedia transmission systems, respectively. For the channel coding, turbo codes with soft output Viterbi algorithm (SOVA) is adopted\cite{wu1999influence}, in which the coding rate is 1/3, the block length is 512, and the number of decoding iterations is 5. 64-QAM is adopted to make the number of transmitted symbols in the traditional communication systems the same as that in DeepSC-S. The details are summarized in Table \ref{speech parameters}.
\renewcommand\arraystretch{1.15} 
\begin{table}[tbp]
\footnotesize
\caption{Parameters settings of the proposed DeepSC-S for telephone systems.}
\label{telephone DeepSC-S NN parameters}
\centering
\begin{tabular}{|c|c|c|c|}
\hline
               & \textbf{Layer Name}  & \textbf{Filters}  & \textbf{Activation}    \\
\hline
    \multirow{2}{6.0em}{\textbf{\centering \ Transmitter}} & 6$\times$SE-ResNet & 6$\times$64   & ReLU \\
\cline{2-4}
                                                & CNN layer   &    8   &   None    \\
\cline{1-4}
    \multirow{3}{6.0em}{\textbf{\centering \ \ \ Receiver}}   & CNN layer   &    8   &   ReLU    \\
\cline{2-4}
                                    & 6$\times$SE-ResNet      &    6$\times$64    &   ReLU    \\
\cline{2-4}
                                    & Last layer (CNN)        &    1              &   None     \\
\hline
\end{tabular}
\end{table}
\renewcommand\arraystretch{1.15} 
\begin{table}[tbp]
\footnotesize
\caption{Parameters settings of the traditional communication systems.}
\label{speech parameters}
\centering
\begin{tabular}{|c|c|c|}
\hline
    & \textbf{Telephone Systems} & \textbf{Multimedia Systems} \\
\cline{1-3} 
    \textbf{Sample rate}     & 8KHz    &   44.1KHz     \\
\cline{1-3}
    \textbf{Samples length} & 16384   &   16384       \\
\cline{1-3}
    \textbf{Number of frames}      & 128     &   128         \\
\cline{1-3}
    \textbf{Frame length}    & 128     &   128         \\
\cline{1-3} 
    \textbf{Source coding}          & 8-bits PCM  &   16-bits PCM     \\ 
\cline{1-3} 
    \textbf{Channel coding}         & Turbo codes  &   Turbo codes     \\ 
\cline{1-3} 
    \textbf{Modulation}             & 64-QAM  &   64-QAM    \\ 
\hline
\end{tabular}
\end{table}

\subsubsection{\textbf{Benchmark 2}}
\begin{figure}[tbp]
\includegraphics[width=0.45\textwidth]{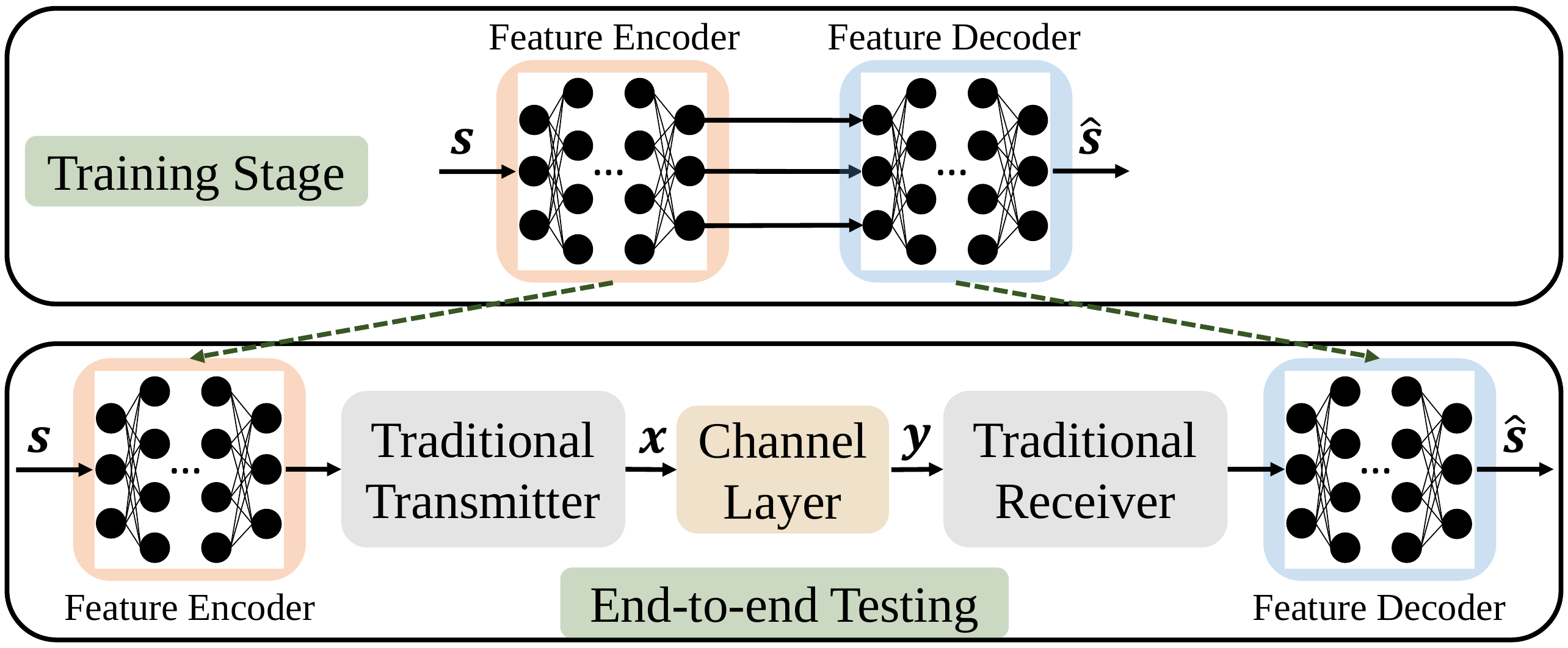} 
\centering 
\caption{The benchmark model by combing a feature encoder with the transmission systems.} 
\label{feature learning}  
\end{figure}

The second benchmark combines feature learning with the traditional channel encoding, named semi-traditional system, as shown in Fig. \ref{feature learning}. At the training stage, the \emph{feature encoder} takes the speech samples, $\boldsymbol s$, as the inputs and its output is fed into the \emph{feature decoder} directly. The received signal is converted into the speech information, $\widehat{\boldsymbol s}$, by the \emph{feature decoder}. Based on signals $\boldsymbol s$ and $\widehat{\boldsymbol s}$, the MSE loss is computed at the end of the receiver, thus, the trainable parameters of the \emph{feature encoder} and the \emph{feature decoder} are updated via SGD at the same time.

For the E2E testing, the pre-trained feature leaning system is split into the \emph{feature encoder} and the \emph{feature decoder}, which are placed before the traditional transmitter and after the traditional receiver, respectively. Note that the signal processing blocks of the traditional transmitter and the traditional receiver are same as the settings as shown in Table \ref{speech parameters}. During the training stage, the \emph{feature encoder} and the \emph{feature decoder} are treated as the extraction and recovery operations without considering communication problems. During the testing stage, the system could yield efficient transmission and mitigate the channel effects. The 2D CNN is adopted in the \emph{feature encoder} and the \emph{feature decoder}. The learning rate is 0.001. The parameters settings are summarized in Table \ref{feature learning system NN parameters}.

\subsubsection{\textbf{Benchmark 3}} 
In order to emphasize the gain from the attention mechanism, we consider a CNN-based semantic communication system without attention mechanism. The learning rate is 0.001. The parameters settings of the CNN-based system are similar to that of the proposed DeepSC-S as shown in Table \ref{telephone DeepSC-S NN parameters}. The only difference is to replace the SE-ResNet modules in the transmitter/receiver for implementing the \emph{semantic encoder/decoder} with the CNN modules. Each CNN module contains 32 \emph{filters}.

\subsection{Complexity Analysis}\label{complexity analysis}
The complexity of the semi-traditional system is higher than that of the traditional one due to the introduction of the \emph{feature encoder/decoder}. Moreover, due to the complexity of NN training, the CNN-based system and the developed DeepSC-S require higher computational cost than the traditional approaches, which can be measured by the floating point operations (FLOPs). For a single 2D CNN module, the required FLOPs can be expressed as\cite{molchanov2016pruning}:
\begin{equation}
{\mathrm{FLOPs}}_{\mathrm{CNN}}=2\times G\times H\times\left(C_{in}\times K^2+1\right)\times C_{out},
\label{cnn flops}
\end{equation}
where $G$ and $H$ are the width and height of the input feature map of the CNN, respectively. $K^2$ is the kernel size. $C_{in}$ and $C_{out}$ are the number of channels\footnote{Here, the channel refers to the parameter of CNN, not the wireless channels.} of the input and output feature maps, respectively.
\renewcommand\arraystretch{1.15} 
\begin{table}[tbp]
\footnotesize
\caption{Parameters settings of the benchmark 2.}
\label{feature learning system NN parameters}
\centering
\begin{tabular}{|c|c|c|c|}
\hline
               & \textbf{Layer Name} & \textbf{Filters} & \textbf{Activation}\\
\hline
    \multirow{2}{7.3em}{\textbf{\centering Feature Encoder}} & 4$\times$CNN modules & 4$\times$32 & ReLU \\
\cline{2-4}
                                            & CNN module   &    8   &   None    \\
\cline{1-4}
    \multirow{3}{7.3em}{\textbf{\centering Feature Decoder}} & CNN module &  8  &   ReLU \\
\cline{2-4}   
                                            & 4$\times$CNN modules & 4$\times$32 & ReLU \\
\cline{2-4}
                                            & CNN module   &    1   &   None    \\
\hline

\end{tabular}
\end{table}

In the semi-traditional system, the number of \emph{filters} of the 5 cascaded CNN modules at the \emph{feature encoder} are 32, 32, 32, 32, 8, respectively. The input size of the \emph{feature encoder} is $128\times128\times1$, i.e., $G=128$ and $H=128$. $K=5$. $C_{in}$ of the 5 CNN modules are 1, 32, 32, 32, 32, respectively, and $C_{out}$ of the 5 CNN modules are 32, 32, 32, 32, 8, respectively. According to (\ref{cnn flops}), the FLOPs required by the \emph{feature encoder} can be calculated as
\begin{equation}
\begin{split}
    \mathrm{FLOPs}_{\mathrm{Semi}}^{\mathcal T}&=2\times128\times128\times\left(1\times5\times5+1\right)\times32 \\
    &+6\times128\times128\times\left(32\times5\times5+1\right)\times32 \\
    &+2\times128\times128\times\left(32\times5\times5+1\right)\times8 \\
    &=2.75\times10^9.
\end{split}
\label{flops feature encoder}
\end{equation}

At the receiver, the input size of the \emph{feature decoder} is $128\times128\times8$ and the number of \emph{filters} of the 6 CNN modules are 8, 32, 32, 32, 32, 1, respectively. Then the FLOPs required by the \emph{feature decoder} are $\mathrm{FLOPs}_{\mathrm{Semi}}^{\mathcal R}=2.81\times10^9$. Therefore, the total FLOPs required by the \emph{feature encoder} and the \emph{feature decoder} are $5.56\times10^9$, i.e., the semi-traditional system requires $5.56\times10^9$ FLOPs more than the traditional system.

In the CNN-based system, the number of \emph{filters} of the 7 CNN modules at the transmitter are 32, 32, 32, 32, 32, 32, 8, respectively. The input size of the CNN-based system is $128\times128\times1$, i.e., $G=128$ and $H=128$. $K=5$. $C_{in}$ of the 7 CNN modules are 1, 32, 32, 32, 32, 32, 32, respectively, and $C_{out}$ of the 7 CNN modules are 32, 32, 32, 32, 32, 32, 8, respectively. According to (\ref{cnn flops}), the FLOPs required by the transmitter are $\mathrm{FLOPs}_{\mathrm{CNN}}^{\mathcal T}=4.44\times10^9$. Similarly, the number of \emph{filters} of the 8 CNN modules at the receiver are 8, 32, 32, 32, 32, 32, 32, 1, respectively. Then the FLOPs required by the receiver are $\mathrm{FLOPs}_{\mathrm{CNN}}^{\mathcal R}=4.49\times10^9$. Therefore, the total FLOPs required by the CNN-based system are $8.93\times10^9$.

Furthermore, in the developed DeepSC-S, a single SE-ResNet module consists of 2 CNN modules with 16 \emph{filters} and kernel size is $5\times5$, as well as 1 CNN module with 32 \emph{filters} and kernel size is $1\times1$. Given the input size of a single SE-ResNet is $128\times128\times32$, so $G=128$ and $H=128$. $C_{in}$ of the 3 CNN modules are 32, 32, 32, respectively, and $C_{out}$ of the 3 CNN modules are 16, 16, 32, respectively. Then the FLOPs required by this SE-ResNet module are ${\mathrm{FLOPs}}_{\mathrm{SE}}=8.75\times10^8$. Accordingly, the total FLOPs required by DeepSC-S are $9.36\times10^9$, including $4.65\times10^9$ FLOPs at the transmitter and $4.71\times10^9$ FLOPs at the receiver, which achieves a $4.82\%$ increase over the CNN-based system.

\subsection{Experiments over Telephone Systems}\label{experiment telephpne}
In this experiment, we investigate a robust system to work under various channel conditions by training DeepSC-S under the fixed channel condition and then testing it under different fading channels. Note that during the training stage, the Gaussian noise contained in the three training channels is generated under a fixed SNR value, 8 dB, because we found that 8 dB is the most suitable value after comparing the models trained under different SNR values.
\begin{figure*}[tbp]
\begin{minipage}[t]{0.33\linewidth}
\centering
\includegraphics[width=1\textwidth]{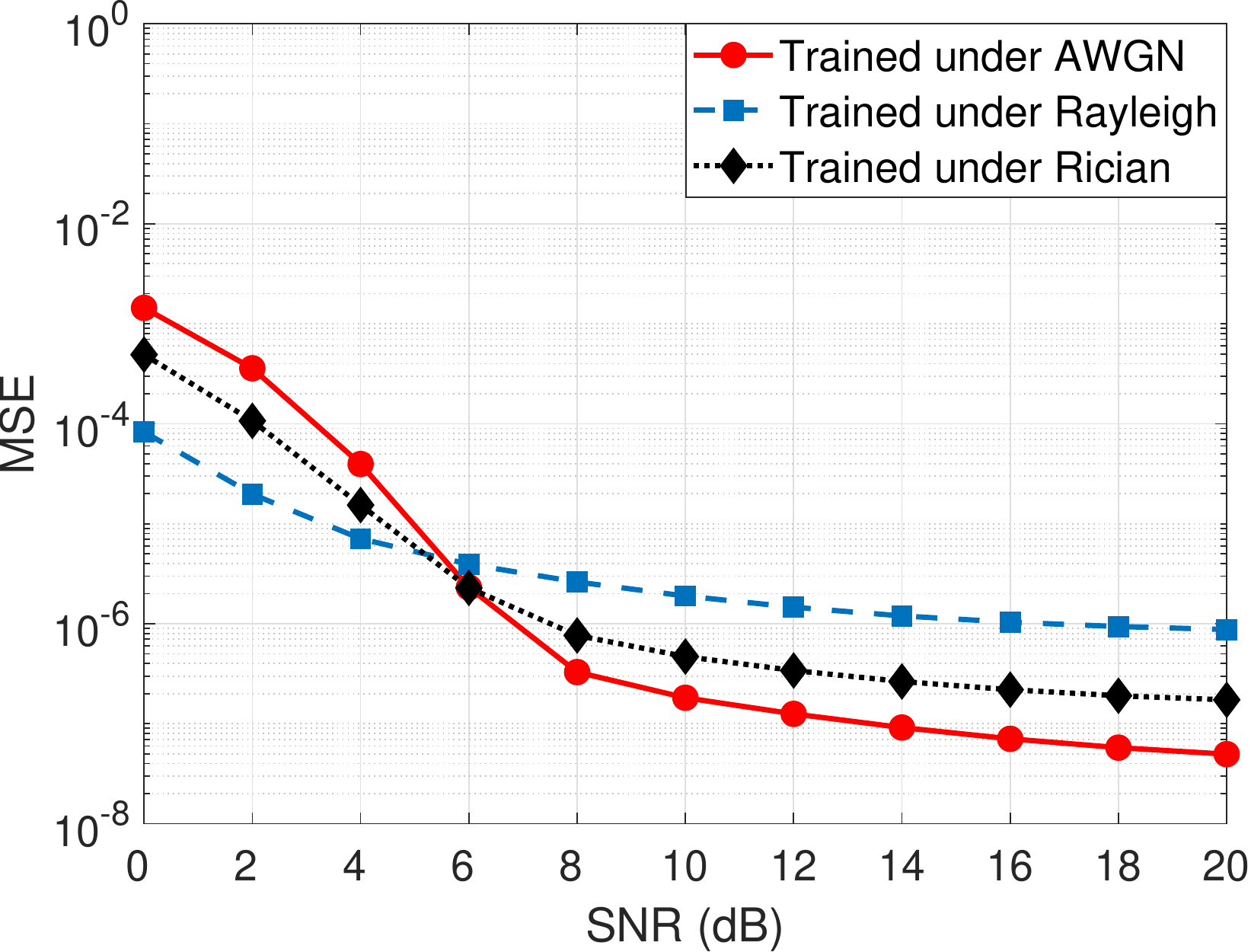} 
\subcaption{AWGN channels}
\label{MSE tested under AWGN channel}
\end{minipage}
\begin{minipage}[t]{0.33\linewidth}
\centering
\includegraphics[width=1\textwidth]{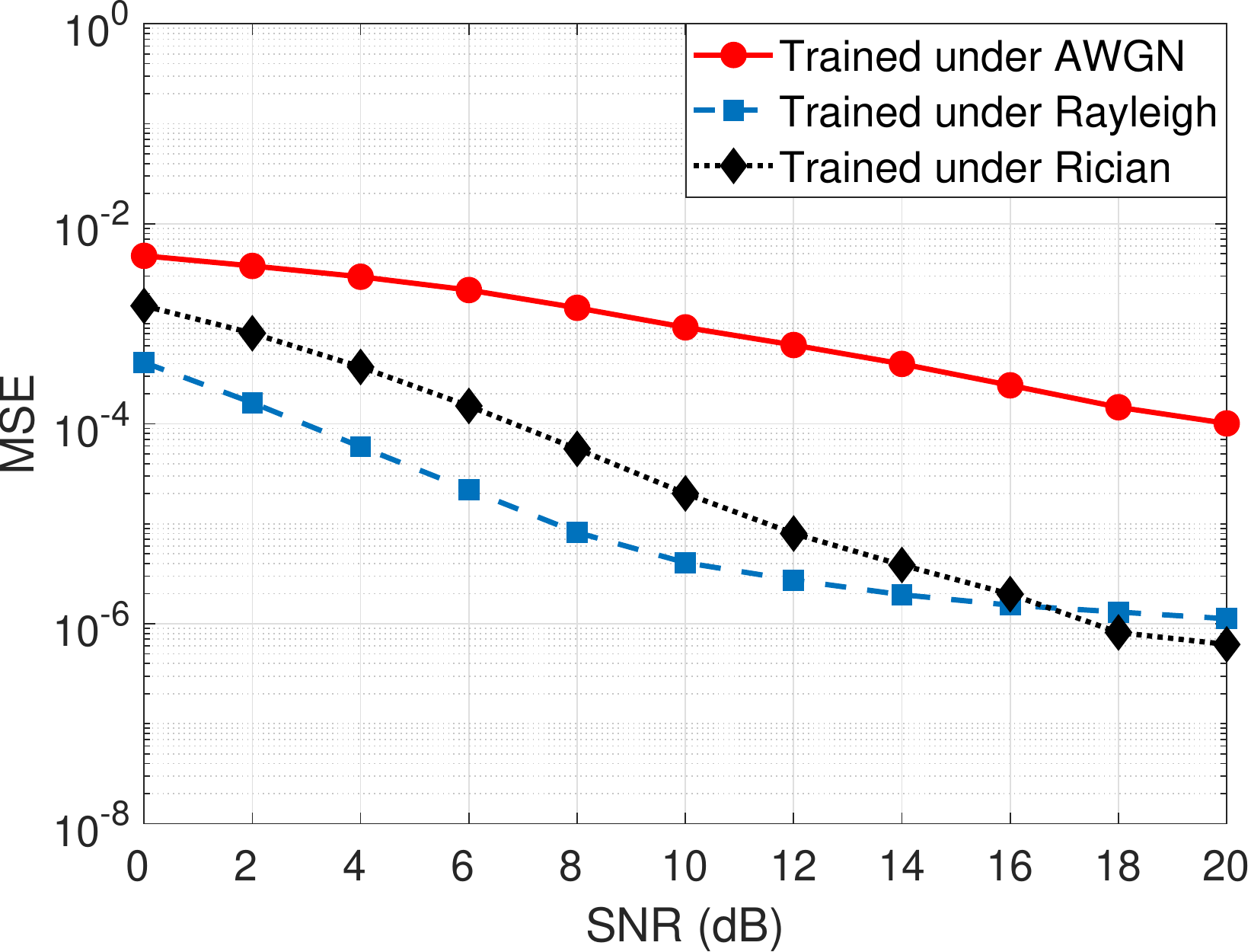} 
\subcaption{Rayleigh channels}
\label{MSE tested under Rayleigh channel}
\end{minipage} 
\begin{minipage}[t]{0.33\linewidth}
\centering
\includegraphics[width=1\textwidth]{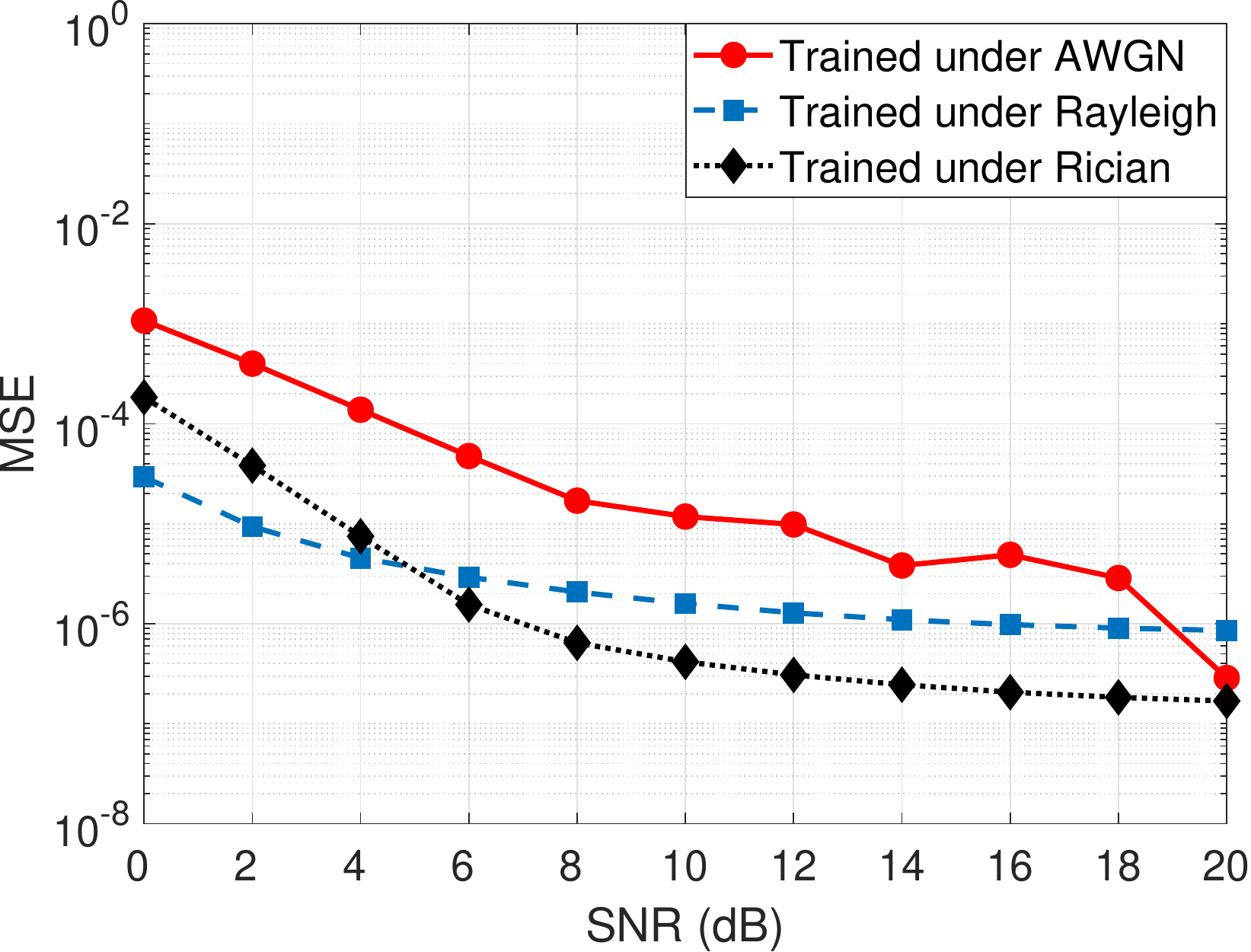}
\subcaption{Rician channels}
\label{MSE tested under Rician channel}
\end{minipage} 
\caption{MSE loss tested under (a) AWGN channels, (b) Rayleigh channels, (c) Rician channels with the models trained under various channels.}
\label{MSE comparison for different testing channels}
\end{figure*}

According to Fig. \ref{MSE comparison for different testing channels}, when SNR is lower than 8 dB, DeepSC-S trained under the AWGN channels has higher MSE loss than the models trained under the Rayleigh channels and the Rician channels. As shown in Fig. \ref{MSE comparison for different testing channels} (a), in terms of the MSE loss tested under the AWGN channels, DeepSC-S trained under the AWGN channels outperforms the model trained under the Rayleigh channels and the Rician channels when SNR is higher than around 6 dB. Besides, according to Fig. \ref{MSE comparison for different testing channels} (b), DeepSC-S trained under the AWGN channels performs quite poor in terms of MSE loss when testing under the Rayleigh channels. Furthermore, Fig. \ref{MSE comparison for different testing channels} (c) shows the model trained under the three adopted channels can achieve MSE loss values under $9\times10^{-7}$ when testing under the Rician channels. Therefore, DeepSC-S trained under the Rician channels is adopted as a robust model that is capable of coping with various channel environments.

\subsubsection{SDR and PESQ Results}
\begin{figure}[htbp]
\includegraphics[width=0.40\textwidth]{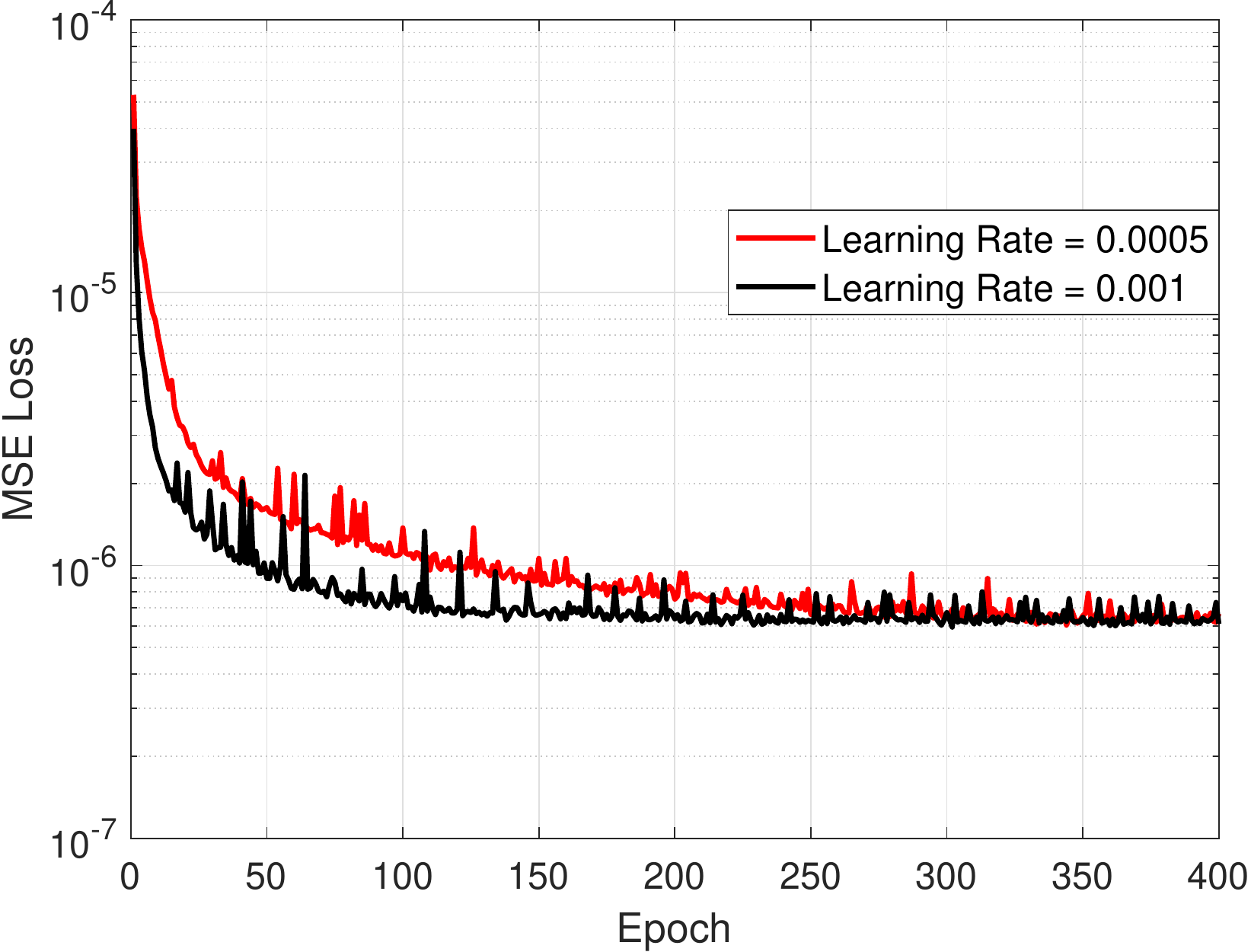}
\centering 
\caption{The training MSE loss versus epoch under the Rician channels with $\mathrm{SNR}=8\;\mathrm{dB}$.}  
\label{loss vs epochs}  
\end{figure}

Based on the roust model, i.e., DeepSC-S trained under the Rician channels with $\mathrm{SNR}=8\;\mathrm{dB}$, the relationship between the MSE loss values and the number of epochs is shown in Fig. \ref{loss vs epochs}. We can observe that the MSE loss reaches convergence after about 400 epochs.

\begin{figure*}[htbp]
\begin{minipage}[t]{0.33\linewidth}
\centering
\includegraphics[width=1\textwidth]{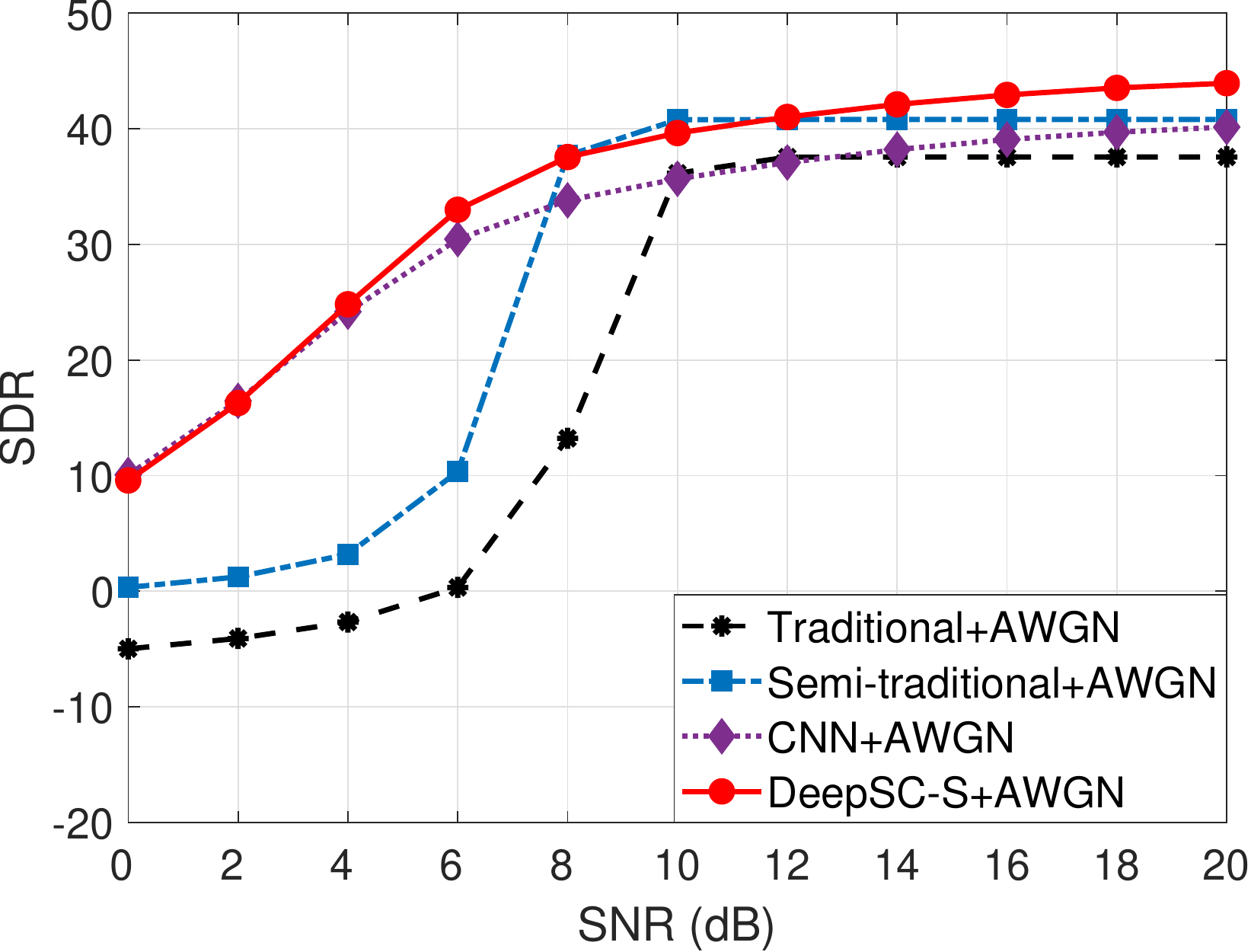} 
\subcaption{AWGN channels}
\label{SDR tested under AWGN channel telephone}
\end{minipage}
\begin{minipage}[t]{0.33\linewidth}
\centering
\includegraphics[width=1\textwidth]{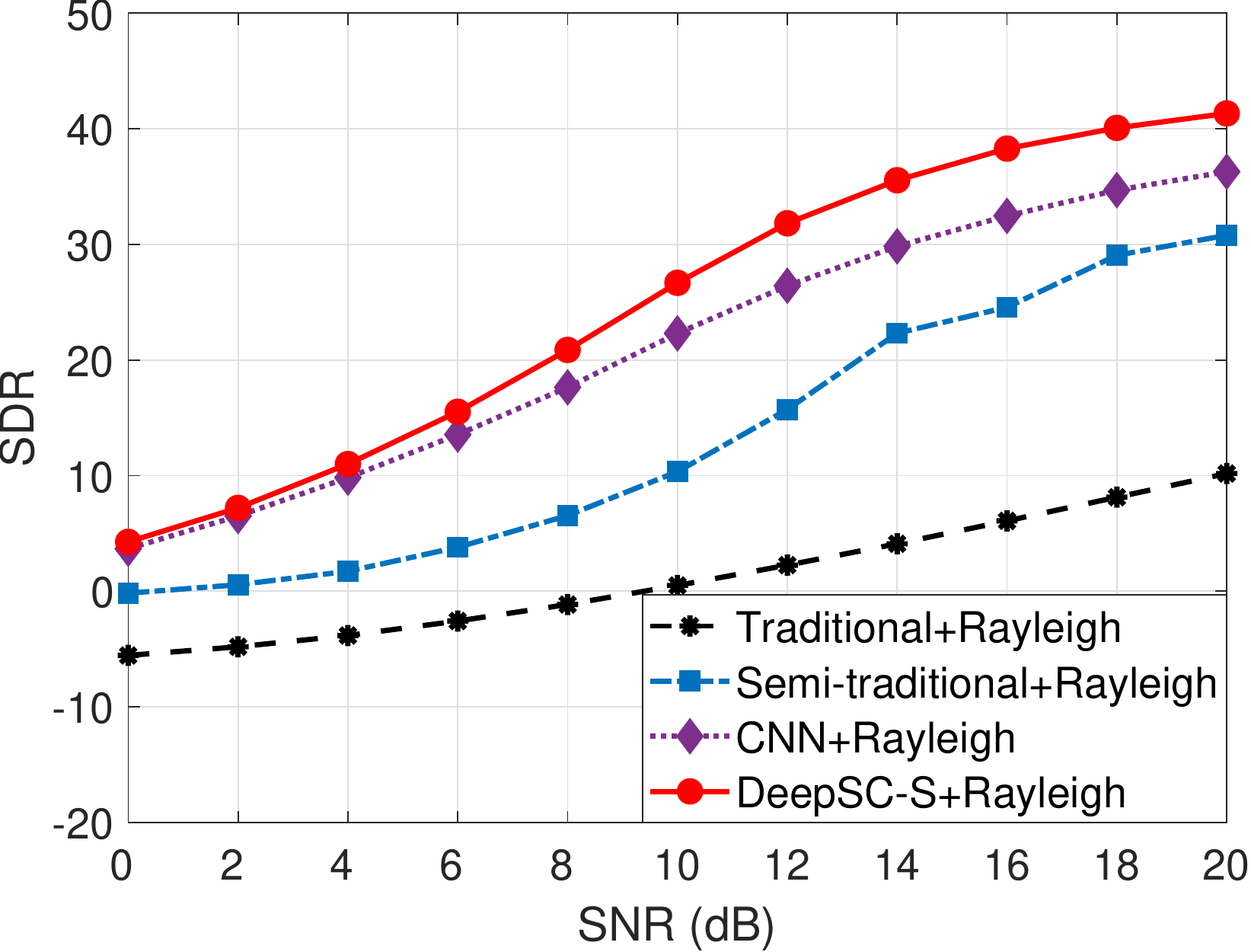} 
\subcaption{Rayleigh channels}
\label{SDR tested under Rayleigh channel telephone}
\end{minipage} 
\begin{minipage}[t]{0.33\linewidth}
\centering
\includegraphics[width=1\textwidth]{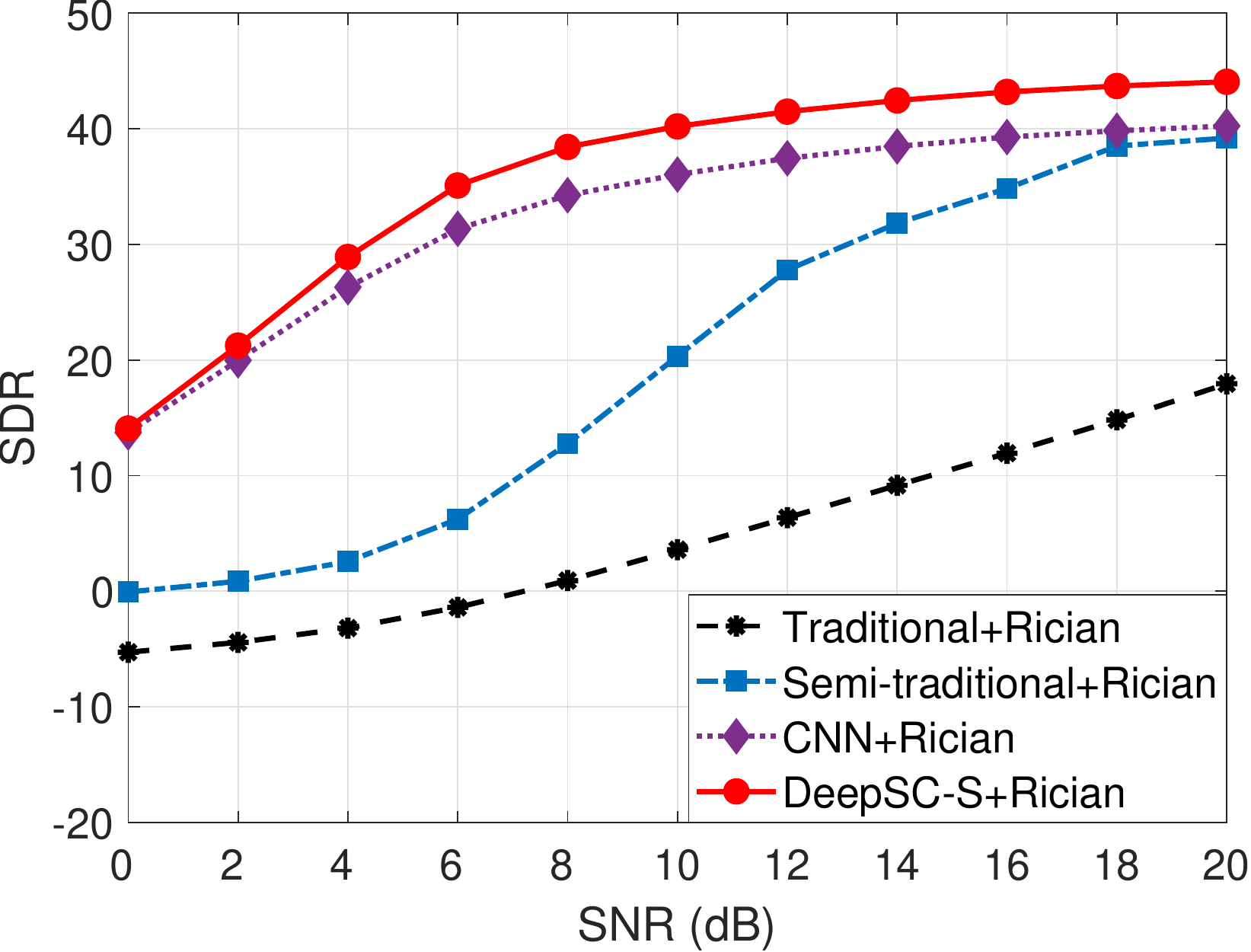}
\subcaption{Rician channels}
\label{SDR tested under Rician channel telephone}
\end{minipage} 
\caption{SDR score versus SNR for speech-based telephone communications with the traditional system, the semi-traditional system, the CNN-based system, and DeepSC-S under (a) AWGN channels, (b) Rayleigh channels, (c) Rician channels.}
\label{SDR result for telephone communcations}
\end{figure*}

Fig. \ref{SDR result for telephone communcations} tests the SDR performance between the three benchmarks and the proposed DeepSC-S under the AWGN, Rayleigh, and Rician channels. From the figure, the semi-traditional system yields higher SDR score than the traditional one under all tested channel environments while its performance is unreliable when SNR is low. Besides, the CNN-based system and DeepSC-S obtain higher SDR score than the other two systems under the Rayleigh channels and the Rician channels, as well as the AWGN channels over most SNR regions. In addition, DeepSC-S performs steadily when coping with different fading channels and SNRs, however, for the semi-traditional system and the traditional system, the performances are quite poor under dynamic channel conditions, especially in the low SNR regime. Moreover, due to the attention mechanism, SE-ResNet, the proposed DeepSC-S achieves higher SDR score than the CNN-based system under all adopted SNRs and fading channels, which proves the effectiveness of DeepSC-S. Particularly, DeepSC-S achieves $10.92\%$ average ascent over the CNN-based system in terms of the SDR performance.

\begin{figure*}[htbp]
\begin{minipage}[t]{0.33\linewidth}
\centering
\includegraphics[width=1\textwidth]{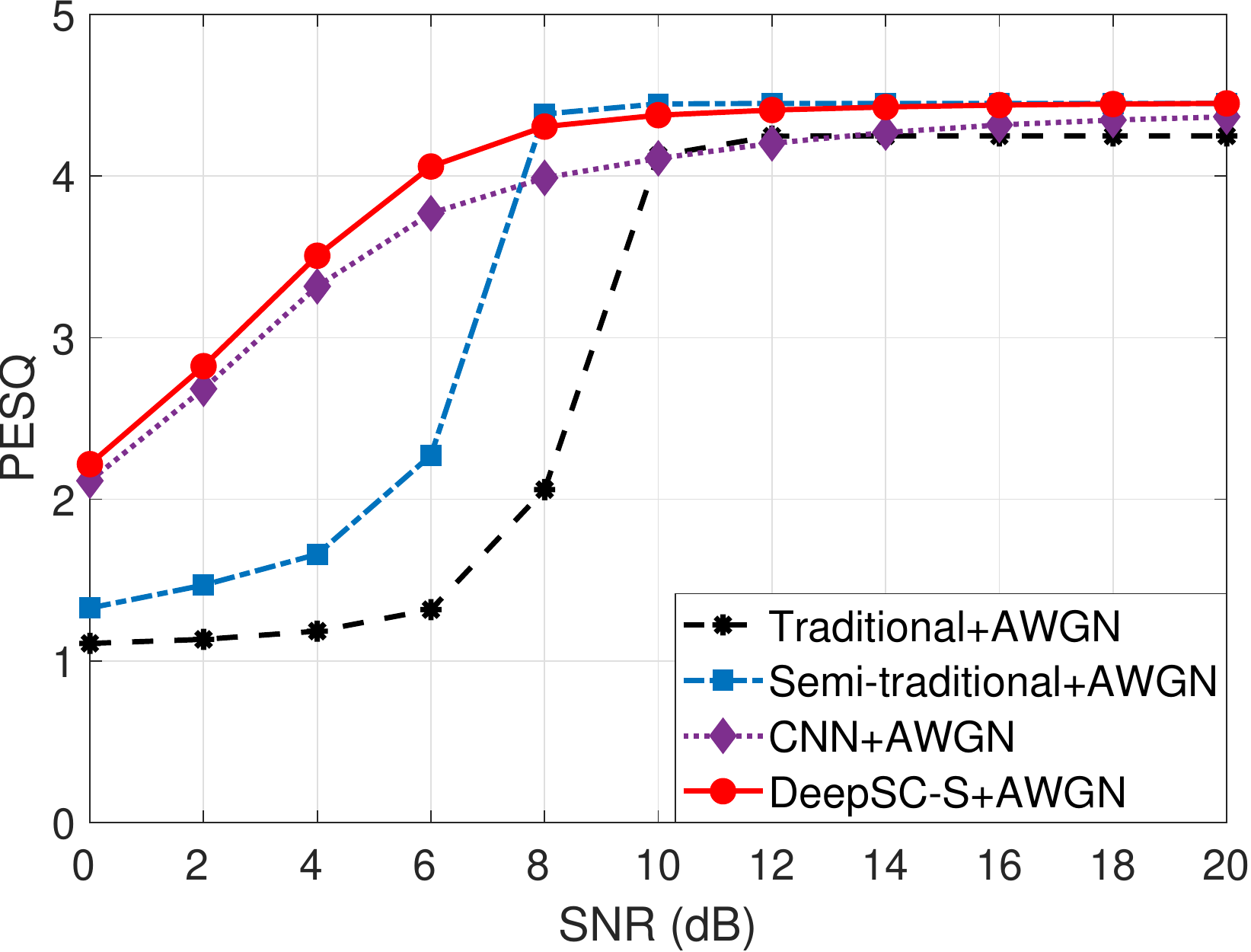} 
\subcaption{AWGN channels}
\label{PESQ tested under AWGN channel telephone}
\end{minipage}
\begin{minipage}[t]{0.33\linewidth}
\centering
\includegraphics[width=1\textwidth]{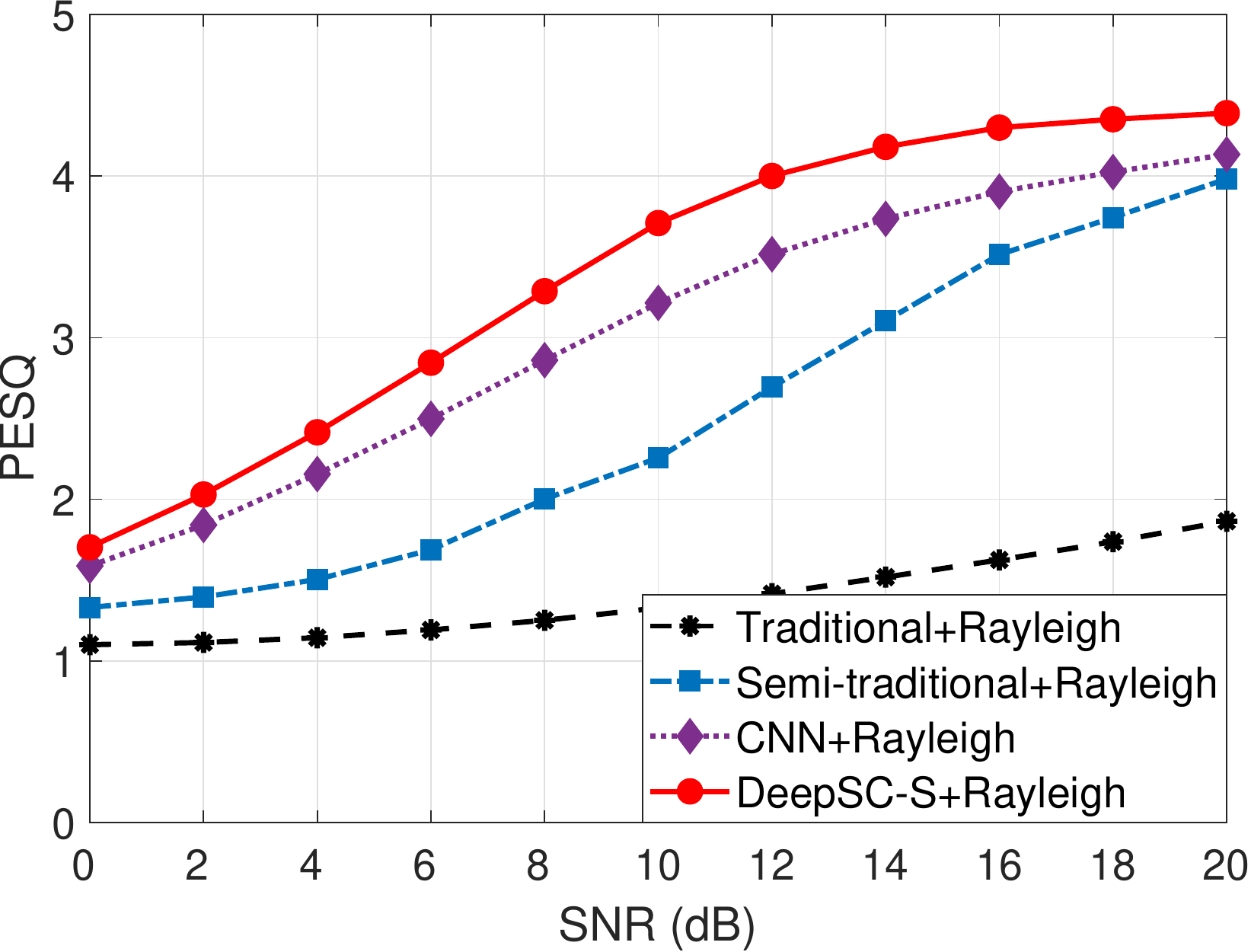} 
\subcaption{Rayleigh channels}
\label{PESQ tested under Rayleigh channel telephone}
\end{minipage} 
\begin{minipage}[t]{0.33\linewidth}
\centering
\includegraphics[width=1\textwidth]{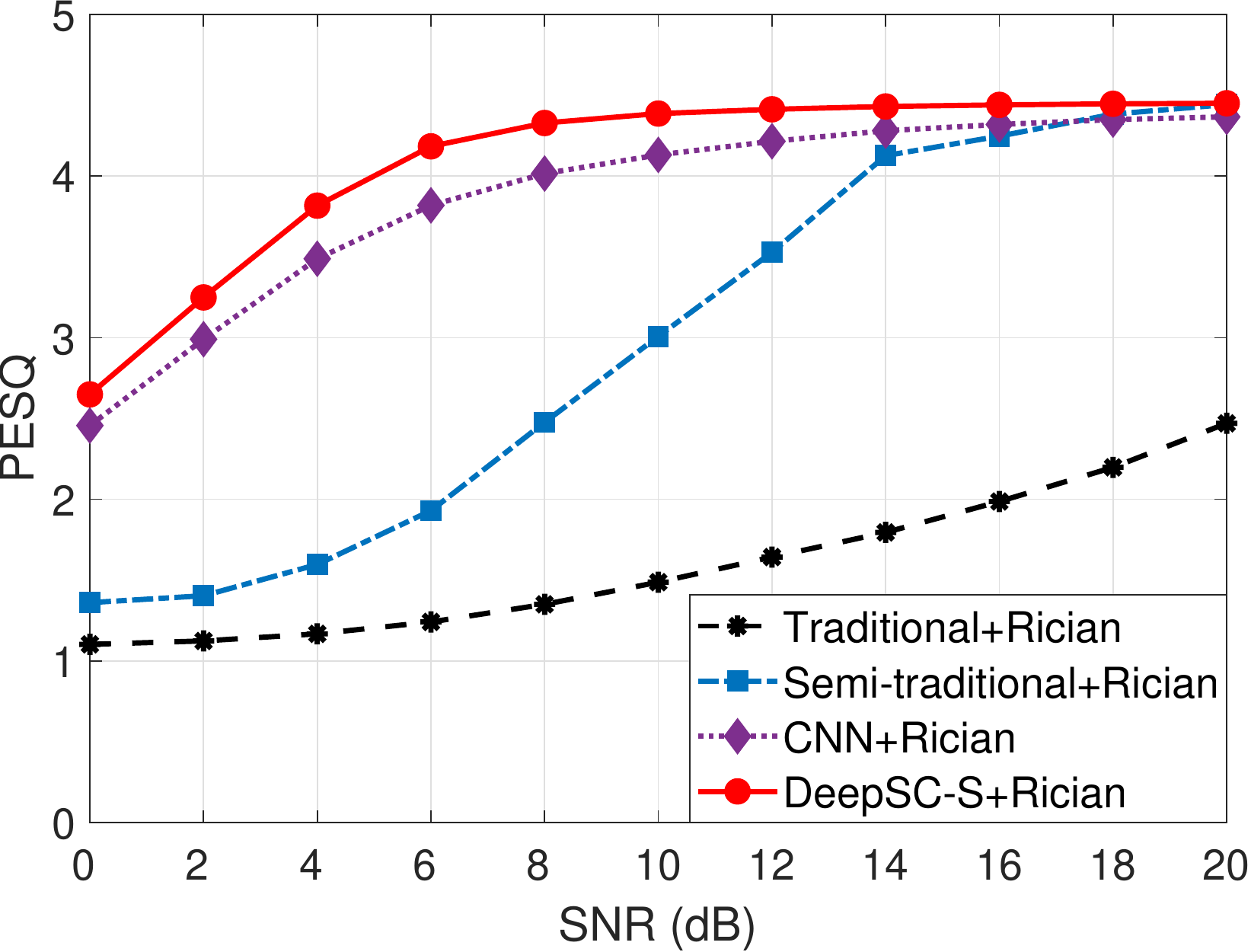}
\subcaption{Rician channels}
\label{PESQ tested under Rician channel telephone}
\end{minipage} 
\caption{PESQ score versus SNR for speech-based telephone communications with the traditional system, the semi-traditional system, the CNN-based system, and DeepSC-S under (a) AWGN channels, (b) Rayleigh channels, (c) Rician channels.}
\label{PESQ result for telephone communcations}
\end{figure*}

The PESQ score comparison is shown in Fig. \ref{PESQ result for telephone communcations}. From the figure, the CNN-based system and DeepSC-S outperform the other two systems under various fading channels and SNRs. Moreover, similar to the results of SDR, DeepSC-S obtains good PESQ when coping with channel variation while the traditional one provides poor scores in the low SNR regime. DeepSC-S also achieves higher score than the CNN-based system under all adopted channel conditions. Particularly, DeepSC-S achieves $7.34\%$ average increase over the CNN-basd system in terms of the PESQ performance. Based on the simulated results, the proposed DeepSC-S is able to yield better speech transmission for telephone systems under complicated communication scenarios than the traditional approaches, especially in the low SNR regime.

\subsection{Experiments over Multimedia Transmission Systems}
In this part, we present the SDR and PESQ performance comparison. The NN parameters settings of the CNN-based system and DeepSC-S are similar to that in the telephone communications experiment, but the number of \emph{filters} of the CNN layer in both systems is 16. Note that the SDR and PESQ results are also tested under the robust model, i.e., DeepSC-S trained under the Rician channels with $\mathrm{SNR}=8\;\mathrm{dB}$.
\begin{figure}[htbp]
\includegraphics[width=0.40\textwidth]{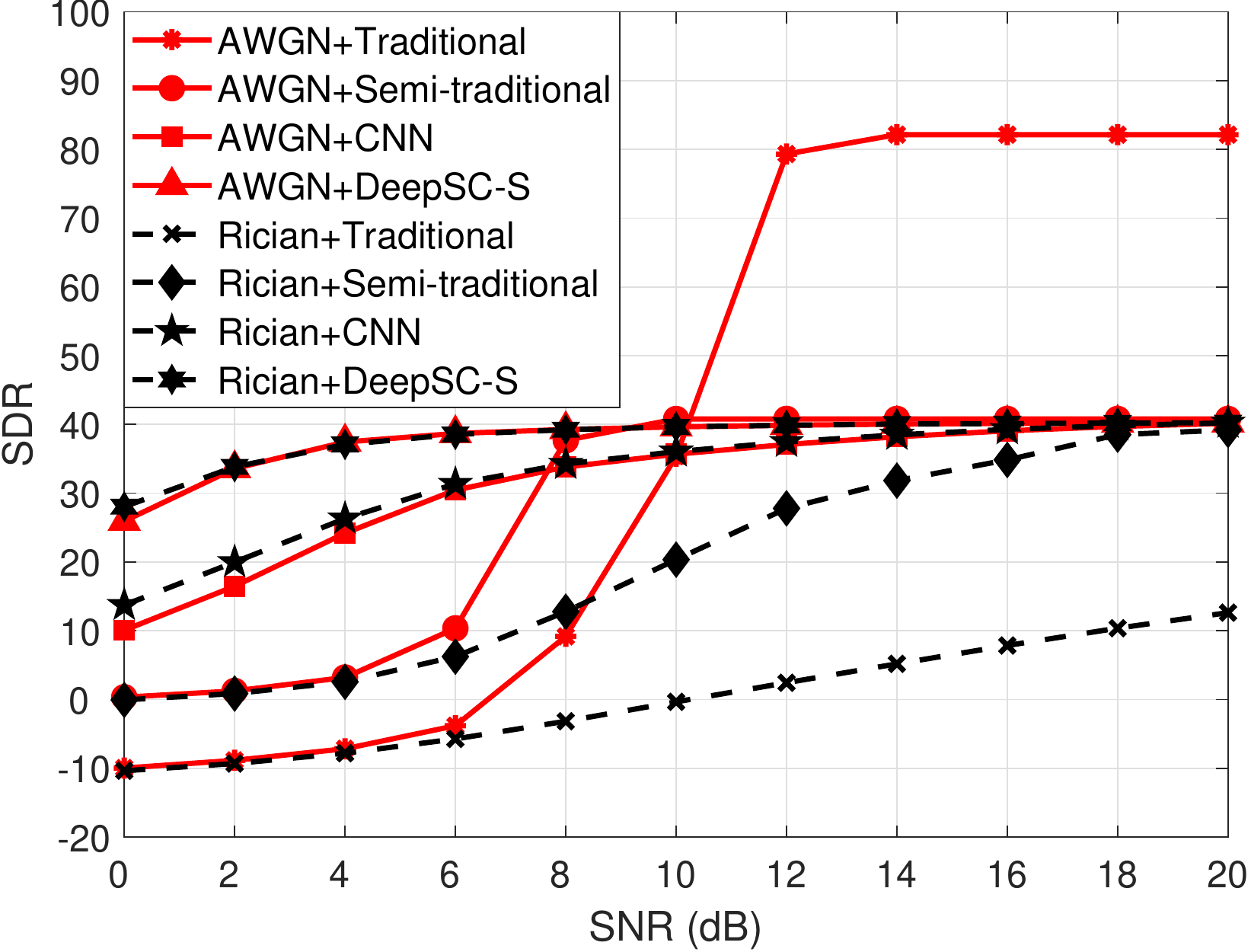}
\centering 
\caption{SDR score versus SNR for speech-based multimedia communications with the traditional system, the semi-traditional system, the CNN-based system, and DeepSC-S under AWGN channels and Rician channels.} 
\label{SDR result for multimedia communcations}  
\end{figure}
\begin{figure}[htbp]
\includegraphics[width=0.40\textwidth]{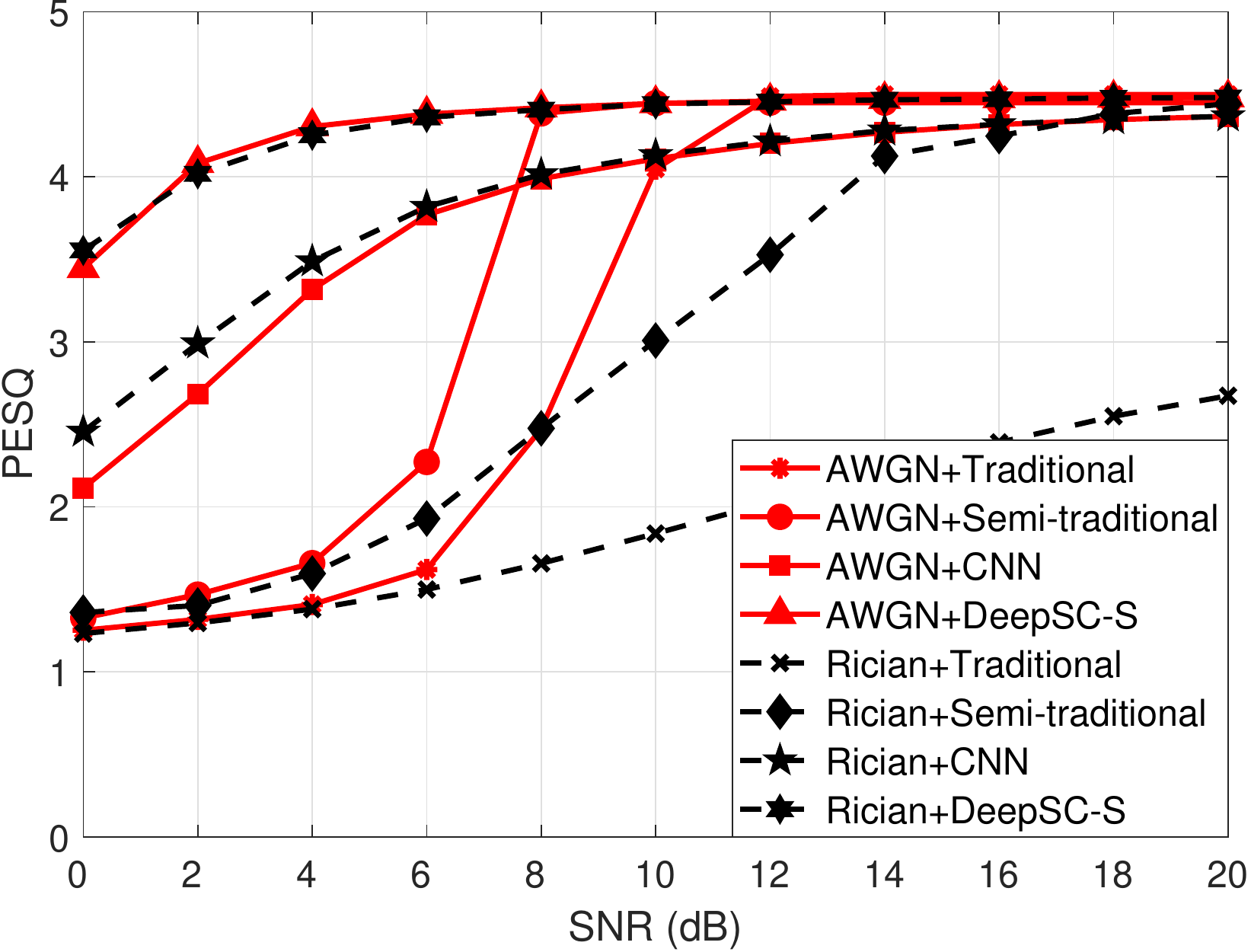} 
\centering 
\caption{PESQ score versus SNR for speech-based multimedia communications with the traditional system, the semi-traditional system, the CNN-based system, and DeepSC-S under AWGN channels and Rician channels.} 
\label{PESQ result for multimedia communcations}  
\end{figure}

Fig. \ref{SDR result for multimedia communcations} depicts the SDR performance comparison for multimedia communications between DeepSC-S and the three benchmarks under the AWGN channels and the Rician channels. Under the AWGN channels, the SDR score of the traditional communication system shows a sharp increase when SNR is higher than 8 dB. Moreover, it achieves SDR scores over 80 in high SNRs due to the high PCM quantization accuracy. However, DeepSC-S can reach universal strong SDR values for all tested SNRs and fading channels. Moreover, DeepSC-S outperforms the semi-traditional system and the traditional system under the Rician channels, as well as the AWGN channels in the low SNR regime. Furthermore, DeepSC-S has higher SDR score than the CNN-based system as the SE-ResNet module is utilized to learn and extract the essential information.

The simulation result of PESQ for multimedia communications is illustrated in Fig. \ref{PESQ result for multimedia communcations}. From the figure, DeepSC-S outperforms the three benchmarks under the Rician channels with any tested SNRs as well as the AWGN channels with low SNR values. Moreover, similar to the results of SDR, the proposed DeepSC-S achieves higher PESQ score than the CNN-based system under all adopted channel conditions. Thus, it is believed that the investigated DeepSC-S is with greater adaptability than the traditional system for speech-based multimedia communications when coping with channel variation.
\section{Conclusion}
In this article, we investigate a DL-enabled semantic communication system for speech transmission, named DeepSC-S, to improve the transmission efficiency by transmitting the semantic information only. Particularly, we jointly design the \emph{semantic encoder/decoder} and the \emph{channel encoder/decoder} to learn and extract the speech features, as well as to mitigate the channel distortion and attenuation for practical communication scenarios. Additionally, an attention mechanism based on a squeeze-and-excitation network is utilized to improve the recovery accuracy by minimizing the mean-square error of speech signals. Moreover, in order to enable DeepSC-S to work well over various physical channels, a DeepSC-S model with strong robustness to channel variations is developed. The adaptability of the proposed DeepSC-S is verified under the telephone systems and the multimedia transmission systems. Simulation results demonstrate that DeepSC-S outperforms the various benchmarks in the low SNR regime. Hence, the proposed DeepSC-S is a promising candidate for speech semantic communication systems.

\bibliographystyle{IEEEtran}
\bibliography{reference.bib}

\end{document}